\documentstyle[graphics]{cupconf}

\ifoldfss
\else
  \ifnfssone
    \newmathalphabet{\mathit}
      \addtoversion{normal}{\mathit}{cmr}{m}{it}
      \addtoversion{bold}{\mathit}{cmr}{bx}{it}
    \newmathalphabet{\mathcal}
      \addtoversion{normal}{\mathcal}{cmsy}{m}{n}
    \else
    \ifnfsstwo
    \fi
  \fi
\fi

\def\gtrsim{\mathrel{\hbox{\rlap{\hbox{\lower4pt\hbox{$\sim$}}}\hbox{$>$}}}}
\def\lesssim{\mathrel{\hbox{\rlap{\hbox{\lower4pt\hbox{$\sim$}}}\hbox{$<$}}}}

\def\hexnumber#1{\ifcase#1 0\or1\or2\or3\or4\or5\or6\or7\or8\or9\or
 A\or B\or C\or D\or E\or F\fi }

\makeatletter
\ifx\CUP@mtlplain@loaded\undefined
\else
\fi
\makeatother
 \makeatletter
 \ifx\CUP@mtlplain@loaded\undefined
   \font\tenbmi=cmmib10 at 10pt
   \font\sevenbmi=cmmib10 at 7pt
   \font\fivebmi=cmmib10 at 5pt

   \newfam\bmifam
   \textfont\bmifam=\tenbmi
   \scriptfont\bmifam=\sevenbmi
   \scriptscriptfont\bmifam=\fivebmi
   
 \fi
 \makeatother
\ifnfsstwo

\fi
\ifnfssone

\fi
\ifoldfss

\fi

\mathchardef\varLambda="0103

\makeatletter
\ifx\CUP@mtlplain@loaded\undefined
\else
\fi
\makeatother
\makeatletter
\ifx\CUP@mtlplain@loaded\undefined
  \font\tenbms=cmbsy10
  \font\sevenbms=cmbsy10 at 7pt
  \font\fivebms=cmbsy10 at 5pt
  \newfam\bmsfam
  \textfont\bmsfam=\tenbms
  \scriptfont\bmsfam=\sevenbms
  \scriptscriptfont\bmsfam=\fivebms

  \edef\bsy@{\hexnumber\bmsfam}
  \mathchardef\bnabla="0\bsy@72
\fi
\makeatother
\title[Element Abundances in Nearby Galaxies]{Element Abundances in Nearby Galaxies}

\author[Donald R. Garnett]%
{D\ls O\ls N\ls A\ls L\ls D\ns R.\ns G\ls A\ls R\ls N\ls E\ls T\ls T\ls 
}

\affiliation{Steward Observatory, University of Arizona, Tucson AZ 85721, USA}

\begin{document}
\ifnfssone
\else
  \ifnfsstwo
  \else
    \ifoldfss
      \let\mathcal\cal
      \let\mathrm\rm
      \let\mathsf\sf
    \fi
  \fi
\fi

\maketitle

\begin{abstract}
In these lectures I present a highly opinionated review of the 
observed patterns of metallicity and 
element abundance ratios in nearby spiral, irregular, and dwarf 
elliptical galaxies, with connection to a number of astrophysical 
issues associated with chemical evolution. I also discuss some of 
the observational and theoretical issues associated with measuring
abundances in H~II regions and  gas and stellar surface densities 
in disk galaxies. Finally, I will outline a few open questions that
deserve attention in future investigations.
\end{abstract}

\firstsection 

\section{Introduction}

The measurement of element abundances in galaxies other than our own
has a roughly forty-year history, beginning with early attempts to 
measure helium abundances in giant H II regions in the Magellanic 
Clouds and M33 (Aller \& Faulkner 1962, Mathis 1962) and 
pioneering studies of heavy element abundances from forbidden lines 
in extragalactic H II regions (e.g. Peimbert \& Spinrad 1970, Searle 
1971, Searle \& Sargent 1972). Since then this field has grown 
tremendously, with high quality oxygen abundance data in some 40 
nearby spiral galaxies and more than 100 irregular and compact dwarf 
galaxies. The amount of data for other elements (C, N, Ne, S, and
Ar) has also improved tremendously, thanks largely to improvements
in visible-wavelength detectors and the launching of spacecraft 
observatories, such as $IUE$, $HST$, and $ISO$, which have opened
up the UV and IR spectral regions for spectroscopy.

The direct importance of determining the distribution of metallicity
and element abundance ratios in galaxies is the contribution these
measurements make to chemical evolution, and by consequence the 
evolution of galaxies. The elements heavier than H and He in stars
and the interstellar medium (ISM) are the accumulated product of 
previous generations of star formation. The overall metallicity 
(usually represented by O/H in H II regions/ISM, and by Fe/H in stars)
is determined by the total amount of previous star formation. Element
abundance ratios, particularly C/O, N/O, or s-process/Fe, track the
relative contributions of low-mass stars and high-mass stars, 
incorporating information on the stellar initial mass function (IMF). 
The abundances can be affected by gas flows (infall, outflow, or 
internal flows). It is possible, with modeling, to infer important
clues to the evolution of galaxies from abundance measurements.

Beyond galaxy evolution, abundance measurements provide important
ancillary information relevant to other very important astrophysical 
problems, including: 
\begin{itemize}
\item The dependence of the I(CO)/N(H$_2$) conversion on environment
and metallicity, which is critical for determining the amount of 
molecular gas in galaxies.
\item The metallicity dependence of the Cepheid period-luminosity
relation, currently under debate with respect to the determination
of the Hubble constant and the extragalactic distance scale.
\item Understanding the color evolution of galaxies. Colors of 
composite stellar populations depend on both age and metallicity;
metallicity measurements thus reduce degeneracies in the interpretation
of colors.
\item Stellar mass loss rates, particularly the radiatively-accelerated
winds of O and Wolf-Rayet stars, likely depend on the metallicity of
the individual stars. 
\item The cooling function of interstellar gas. Cooling of interstellar
gas is generally dominated by metals (ions of metals in X-ray and
ionized gas, singly-ionized metals in neutral gas, and molecules
other than H$_2$ in molecular clouds), so the thermal balance in
the ISM is a function of metallicity, with obvious implications for 
the formation of stars and galaxies.
\end{itemize}

In any field of investigation, a few key questions arise which form
a framework for specific studies. I formulate a few of them below.
\begin{enumerate}
\item How do metallicity and element abundance ratios evolve within
galaxies, and how do variations relate to the evolution of the 
gas content and stellar light?
\item What galaxy properties determine the observed compositions of
galaxies? How is metallicity affected by galaxy dynamics (interactions,
gas flows, angular momentum evolution)?
\item How did heavy elements get into the intergalactic medium (IGM)?
Were they ejected from galaxies by supernova-driven winds, ejected
in tidal streams during galaxy interactions and mergers, or did they
come from the first, possibly pre-galactic, stars?
\item How well do simulations of galaxy formation and evolution 
reproduce the observed metallicities and distribution of abundances 
in galaxies?
\end{enumerate}

The purpose of these lectures is to review the results of a variety
of element abundance studies in galaxies other than our own in the
nearby universe. I will not try to be all-inclusive, as the field is
vast. For example, I will not attempt to discuss abundances in 
elliptical galaxies in detail, as better experts have already written
extensive reviews on the subject (e.g. Worthey 1998, Henry \& Worthey 
1999), nor will I say much about luminous IR starbursts or low surface
brightness galaxies. Much of the methodology behind abundance measurements 
in stars and ionized gas will be covered 
in great detail in the lectures by Lambert, Langer, and Stasi\'nska; I 
will not spend much time on these subjects, but will highlight points 
of contention or uncertainty where appropriate. Likewise, Matteucci 
will discuss chemical evolution modeling in detail, so I will use the 
observational results to highlight areas where the data shed light on 
physical evolution of galaxies.

\section{Observational Methods for Measuring Abundances}

\subsection{Spectroscopy of H II Regions and Planetary Nebulae}

Most of the information we have on abundances in spiral and irregular
galaxies have come from spectroscopy of H II regions. This is logical 
since H~II regions are luminous and have high surface brightness (in 
the emission lines) compared to individual stars in galaxies. One can
think of an H~II region as an efficient machine for converting the
extreme ultraviolet radiation of a hot, massive star into a few narrow 
emission lines, leading to a very luminous object in the optical/IR
bands. As a result, 
observations of H~II regions have typically provided our first look at 
abundances within galaxies. Indeed, emission lines are now being used
to probe the ISM of galaxies at redshifts greater than 2, as will be
discussed by Pettini in these proceedings.

Elements that are readily observed in the visible spectrum of H~II
regions include O, N, Ne, S, and Ar. With the exception of O, all of
these elements may have important ionization states that emit only
in the ultraviolet or infrared (for example, Ne$^+$, N$^{+2}$, S$^{+3}$).
If we add UV spectroscopy we can study C and Si. Figure 1 shows an HST 
spectrum of one H~II region in the SMC, showing the rich variety of 
forbidden emission lines and H, He recombination lines in the UV and 
optical spectrum. Other abundant heavy 
elements (such as Fe or Mg) may be observed in photoionized nebulae. 
One must always keep in mind that many elements in the ISM are strongly
depleted onto grains, which affects the total abundance. This is an
important but poorly known factor in many cases (such as O and C). Another
caution is that H~II regions show the composition of the present-day
ISM, and are insensitive to the evolution of abundances with time.

\begin{figure}[t!]
\resizebox{\hsize}{!}{\includegraphics{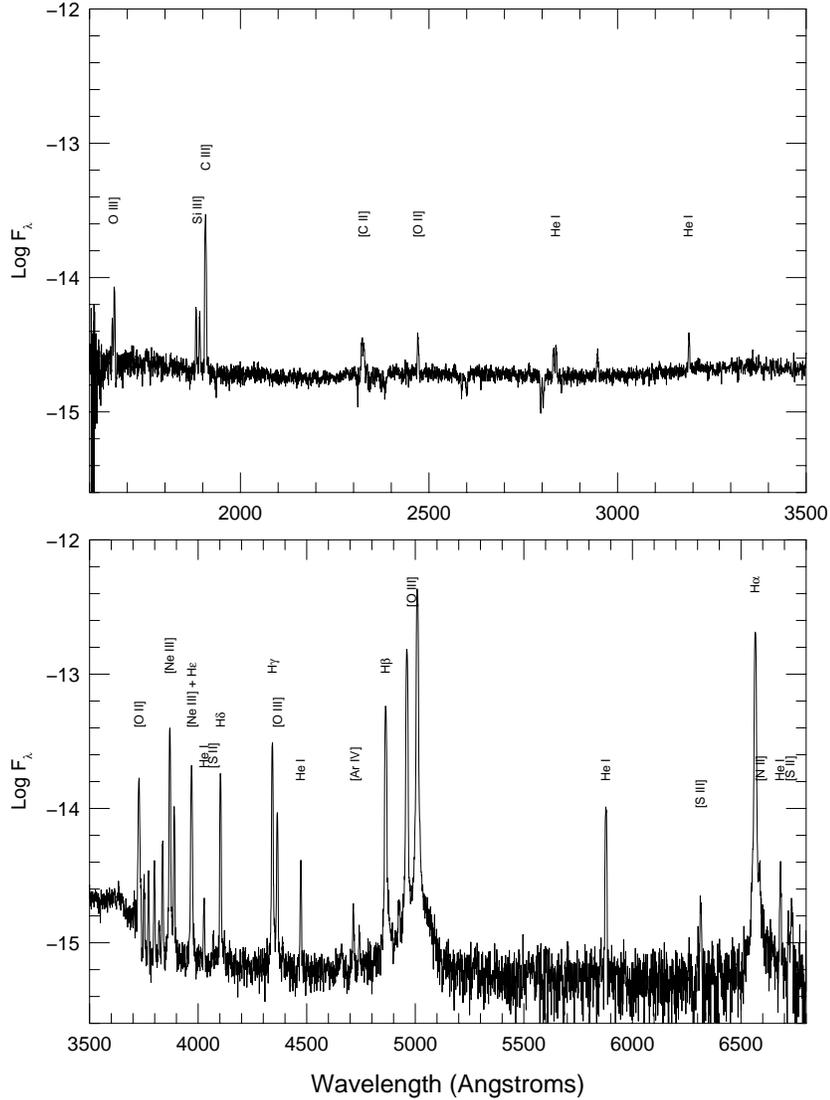}}
\hfill
\parbox[b]{\hsize}{
\label{fig_n88hst} 
\caption{{\it Hubble Space Telescope} UV/optical spectrum of the H~II region
N88A in the Small Magellanic Cloud.}}
\end{figure}

Planetary nebulae (PNs) are essentially H~II regions created by the 
ionization of a red giant envelope by the exposed hot stellar core, 
so spectroscopy of a PN provides information on a similar variety of
elements as H~II regions, with the same caveats. There are a number
of significant differences, however. PNs are much less luminous than
H~II regions, so observations suitable for abundance measurements are
restricted to the nearest galaxies; at the present time, measuring 
abundances in PNs is challenging in galaxies as close as M31 (Jacoby
\& Ciardullo 1999). Another difference is that the PN abundances are altered
from the original stellar composition by nucleosynthesis -- He, C, and
N are often enriched, and even O may be affected. Thus, the use of PNs
to measure abundances across galaxies must be pursued with caution. 
Nevertheless, measurements of PNs offer a means of measuring abundances
across galaxies and their evolution over the range of ages of PN 
progenitors (a few tens to a few thousands of Myr). Stasi\'nska will
discuss PN abundance measurements in her lectures. 

The observational and analytic techniques for determining abundances
in H~II regions have been discussed at great length by Skillman (1998)
at the VIIIth Canary Island Winter School and by Stasi\'nska in her
lectures here. I will add a few remarks here on observing and deriving
abundances.

\subsubsection{Observational Considerations}

It is easy to obtain a high-quality spectrum of an H~II region in a
nearby galaxy. It is not so easy to obtain a high-quality analysis 
afterward. Photon statistics is not the entire story in CCD spectroscopy. 
Additional random errors creep in during the flat-fielding and photometric 
calibration stages. It is difficult to flat-field a CCD frame to better 
than 1\% even in imaging observations, where the most precise flat-fields
involve matching the color of the target to that of the flat-field source
(the night-sky for deep imaging -- see Tyson 1986). Spectroscopists rarely
observe such practices. In typical H~II
region spectroscopy, the flat-field is often obtained by combining an
internal lamp to map the pixel-to-pixel sensitivity variations with a
twilight sky observation to fit the slit vignetting. Both fill the slit
in a different way than the object, which is an important consideration
for the correction of interference fringing in the red. Flat-fields
$repeatable$ to 1\% precision can be obtained over limited areas of a
CCD spectrum, but the precision can be worse over regions where the
lamp source is weak (in the blue part of the spectrum for example) or
vignetting is strong.

The photometric
calibration also contributes to the uncertainty of the measured spectrum.
Flux standard stars are typically measured at widely spaced wavelengths
(50 \AA\ is common), and the sensitivity function of the instrument is
determined by fitting a low-order polynomial or spline to the flux points.
Such fits inevitably introduce low-order ``wiggles'' in the sensitivity
function, which will vary from star to star. Based on experience, the best
spectrophotometric calibration yield uncertainties in the $relative$ 
fluxes of order 2-3\% for widely-spaced emission lines; the errors may
be better for ratios of lines closer than 20 \AA\ apart. Absolute fluxes 
have much higher uncertainties, of course, especially for narrow-aperture 
observations of extended objects. 

Another source of concern is the extended nature of H~II regions and 
patchiness in interstellar reddening, which affects the measured line 
ratios. H~II region spectra are often presented as integrated over 
the source. Reddening by dust is patchy everywhere we look, so the
effects on the H~II region spectrum must vary from point-to-point if
we look at the spatial distribution. Although the spectrum of an H~II
region may be dominated by the areas with the highest surface brightness, 
it may be possible for a bright but obscured area to be given low weight, 
or for a region with a lower-quality spectrum to have an inflated
surface brightness because of poorly-measured extinction. Thus the
patchy nature of dust reddening must introduce additional uncertainty
into measured line ratios. Spatially-resolved measurements are encouraged
whenever possible.

The highly opinionated point here is that anyone who presents measured 
emission line strengths with uncertainties of 1\% or less is probably
not adding in all the error sources. Five percent uncertainties are 
probably more realistic for the brightest emission lines observed.
Note that this level of precision is more than adequate for abundance 
measurements for most astrophysical problems.

\subsubsection{The Direct Method }

Direct abundance measurements can be made when one is able to
measure the faint emission lines which are important diagnostics
of electron temperature, $T_e$. 
The abundance of any ion relative to H$^+$ derived from the ratio of
the intensity of a transition $\lambda$ to the intensity of H$\beta$ 
is given by
\begin{equation}
{N(X^{+i})\over N(H^+)} = {I(\lambda)\over I(H\beta)}{\epsilon(H\beta)\over
\epsilon(\lambda)},
\end{equation}
where $\epsilon(\lambda)$ represents the volume emission coefficient
for a given emission line $\lambda$. For collisionally-excited lines in 
the low-density limit, the analysis in section 5.9 of Osterbrock (1989) 
applies. 

When $T_e$ has been measured, the volume emission coefficient for a 
collisionally-excited line is given by
\begin{equation}
\epsilon(\lambda) = h\nu q_{coll}(\lambda) = {hc\over \lambda}~8.63\times10^{-6}(\Omega/\omega_1)T_e^{-0.5}e^{-\chi/kT_e} 
\end{equation}
where $\Omega$ is the collision strength for the transition observed,
$\omega_1$ is the statistical weight of the lower level, and $\chi$ is 
the excitation energy of the upper level. $\Omega$ contains the physics
in the calculation; it represents the electron-ion collision cross-section
averaged over a Maxwellian distribution of electron velocities relative
to the target ion at the relevant temperature. Thus $\Omega$ has a mild 
temperature dependence, which can introduce a trend in abundance ratios
if not accounted for. 

{\it Note on collision strengths: the vast majority of these values are 
computed, not experimental. This does not mean that they have zero 
uncertainty!} A recent example is given by the case of [S~III] (Tayal \& 
Gupta 1999). This new 27-state R-matrix calculation resulted in changes
of approximately 30\% in the collision strengths for optical and IR 
forbidden transitions from earlier calculations. This shows that even
for commonly-observed ions the atomic data is still in a state of flux.
Observers should take into account the probable uncertainty in atomic
data when estimating errors in abundances.

\begin{figure}[t!]
\resizebox{\hsize}{!}{\includegraphics{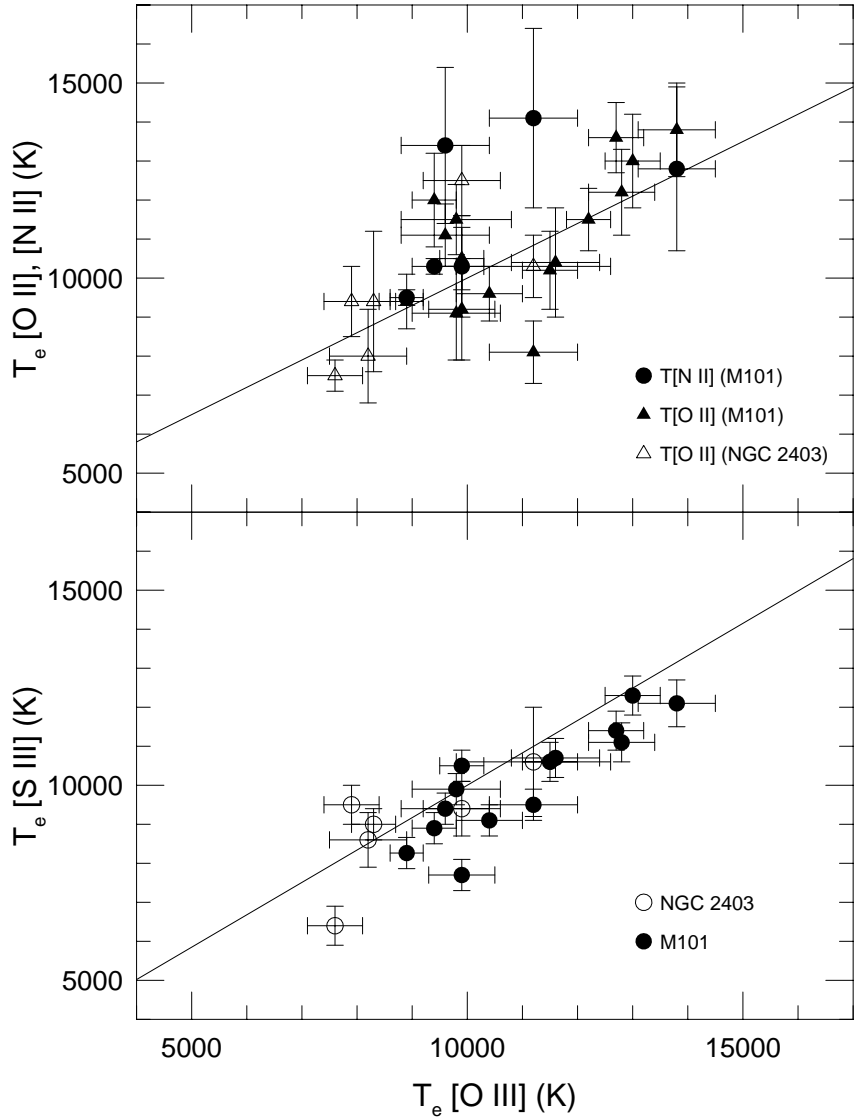}}
\hfill
\parbox[b]{\hsize}{
\label{fig_TCOMP} 
\caption{Comparison of electron temperatures derived from [O~III],
[O~II], [N~II], and [S~III] measurements for H~II regions in NGC 2403
and M101. The straight lines show the correlations predicted by 
photoionization models (Garnett 1992).
}}
\end{figure}

Another thing to account for is the fact that ionized nebulae are not
strictly isothermal. Because [O~III] is usually the most efficient
coolant, the thermal balance at any point in an H~II region depends on 
the local abundance of O$^{+2}$, as well as the local radiation field.
The {\it ion-weighted} electron temperature for a given ion can vary 
with respect to T(O~III) in a predictable way (Garnett 1992), depending
largely on the metallicity. Figure 2 shows a plot of measured electron 
temperatures for [O~III], [S~III], [O~II], and [N~II] compared with
the relationships derived from model photoionized nebulae (solid lines).
The measured temperatures show correlations which agree quite well with
the model relations, although there is quite a bit of scatter in the
[O~II] temperatures, and there may be a slight offset between T[S~III]
and the predicted relation, which may be real or an observational artifact. 
These results indicate that the photoionization models provide a reliable 
predictor of the thermal properties of H~II regions. 

For recombination lines, the emission coefficient is given by
\begin{equation}
\epsilon(\lambda) = h\nu q_{rec}(\lambda) = {hc\over \lambda}~\alpha_{eff}(\lambda),
\end{equation}
where $\alpha_{eff}$($\lambda$) is the ``effective'' recombination coefficient
for the recombination line $\lambda$. $\alpha_{eff}$ incorporates the physics,
including the cross-section for electron-ion recombination and the probability
that a given recombination will produce the given emission line. $\alpha$ values
for H vary as roughly $T_e^{-1}$; individual lines have mildly different 
$T$ dependences, but recombination line ratios are only weakly dependent on 
$T$, and quite insensitive to $n_e$ for densities less than 10$^6$ cm$^{-3}$.

Most astronomers are familiar with the bright H~I Balmer and He~I recombination 
lines in the optical spectrum of ionized nebulae. Heavier elements also emit
a recombination spectrum, and O~I, O~II, C~II, N~I, N~II and other permitted
lines have been observed in PNs and the Orion Nebula. In principle, such 
recombination lines could yield more accurate abundances than the forbidden
lines, because their emissivities all have roughly the same $T$ dependence.
In practice, the recombination lines scale roughly with element abundance,
so even for O and C the RLs are typically fainter than 1\% of H$\beta$,
making them too faint to observe routinely in extragalactic H~II regions.
It is observed that recombination lines in some PNs give much higher abundances
than the corresponding forbidden lines from the same ions (Liu et al. 1995, 
2000; Garnett \& Dinerstein 2001, 2002), and there is currently a raging 
debate over whether the recombination lines or the the forbidden lines provide 
more reliable abundances. 

Measurements of infrared collisionally-excited fine-structure lines
are gaining ground with the launch of the $ISO$ spacecraft, and with
the upcoming $SIRTF$, $SOFIA$, and $FIRST$ missions. Recognizing that 
$\chi$ $\approx$ 5-10 eV for UV forbidden lines, $\chi$ $\approx$ 2-3 eV 
for optical forbidden lines, and $\chi$ $<$ 0.2 eV for IR fine structure 
lines with $\lambda$ $>$ 7$\mu$m, we see that the exponential term 
in Equation 2.2 goes to nearly unity, and the IR lines have a weak
temperature dependence. Thus it should be possible to determine 
accurate abundances free of concerns over temperature fluctuations.
One caveat is that the very important [O~III] and [N~III] fine-structure
lines are sensitive to density, suffering from collisional de-excitation
at $n_e$ $\approx$ 1000 cm$^{-3}$, so density fluctuations could 
introduce large uncertainties. Fine structure lines from Ne, S, and Ar
in the 7-20$\mu$m range, however, are not so sensitive to density. 

For extragalactic H~II regions, the main limitations on IR observations
so far have been small telescopes, high background, and short spacecraft 
lifetimes. Nevertheless, $ISO$ is providing some information on H~II 
regions in the Galaxy and other Local Group galaxies (and luminous 
starbursts), and the future missions promise even better data. 

\subsubsection{``Empirical'' (Strong-Line) Calibrations}

In many cases $T_e$ can not be measured, either because the nebula
is too faint or it is so cool that the temeprature-sensitive diagnostic
lines (for example [O~III] $\lambda$4363) are too weak. Thus, there
is interest in having an abundance indicator that uses the strong
forbidden lines. 

Pagel et al. (1979) identified the line intensity ratio 

\begin{equation}
R_{23} = {{I([O~II]\lambda3727) + I([O~III]\lambda\lambda4959,5007)} \over
{H\beta}}
\end{equation}

\noindent
as an indicator of O/H in H~II regions. They noted, based on a sample of
extragalactic H~II regions, that the measured $T_e$, O/H, and $R_{23}$
were all correlated. This works because of the relationship between O/H
and nebular cooling: the cooling in the ionized gas is dominated by emission
in IR fine-structure lines (primarily the [O~III] 52$\mu$m and 88$\mu$m
lines), so as O/H increases, the nebula becomes cooler. In response, the
optical forbidden lines, especially the [O~III] lines, become weaker as
O/H increases (excitation goes down as $T$ decreases). 

The $R_{23}$ vs. O/H relation is fairly well calibrated empirically (based
on abundances using the direct method) for log O/H between $-$3.5 and $-$4.0
(Edmunds \& Pagel 1984). For higher O/H, the strong-line method breaks down
because few measurements of $T_e$ exist; only two measurements have been
made for H~II regions with roughly solar O/H (Kinkel \& Rosa 1994; Castellanos
et al. 2001). In this regime, the relation has been calibrated using 
photoionization models (which I'll discuss later) that may have systematic
errors. One other complication is that for log O/H $<$ $-$3.8, the relation
between $R_{23}$ and O/H reverses, such that $R_{23}$ decreases with
decreasing abundance. The relation thus becomes double-valued, and at the
turn-around region the uncertainties in O/H are much larger. This occurs
because at very low metallicities the IR fine-structure lines no longer 
dominate the cooling because there are too few heavy elements. As a result 
the forbidden lines more directly reflect the abundances in the gas. 

This double-valued nature of $R_{23}$ has led some to seek other strong-line
diagnostics. The ratio [O~III]/[N~II] (Alloin et al. 1979; Edmunds \& Pagel
1984) has been promoted to break the degeneracy in $R_{23}$. This ratio
does appear to vary monotonically with O/H, although the observational
scatter generally is larger than for $R_{23}$. More recently, the emission
line ratio 

\begin{equation}
S_{23} = {{I([S~II]\lambda\lambda6717,6731) + I([S~III]\lambda\lambda9069,9532)} 
\over {H\beta}}
\end{equation}

\noindent
has been calibrated as an indicator of O/H by D\'\i az \& P\'erez-Montero
(2000). $S_{23}$ has the advantage of varying monotically over the range
$-$4.3 $<$ log O/H $<$ $-$3.7 in which $R_{23}$ becomes ambiguous. $S_{23}$
does become double-valued for O/H $>$ $-$3.4. Where this relation breaks
down is uncertain at present because there are too few measurements. In
addition, the ratio [N~II]$\lambda$6583/H$\alpha$ has been promoted as 
another possible measure of O/H (van Zee et al. 1998; Denicol\'o, Terlevich 
\& Terlevich 2001). [N~II]/H$\alpha$ varies monotonically with O/H over the
entire range over which it is calibrated, but the scatter is quite large,
especially at low values of O/H in dwarf irregular galaxies. Note that 
$S_{23}$ and [N~II]/H$\alpha$ are employed here as measures of the $oxygen$ 
abundance, not sulfur or nitrogen and are calibrated by direct measurements 
of O/H. Thus, non-solar abundance ratios are not a concern. 

At the same time, there are several limitations. 
\begin{enumerate}
\item None of these strong-line diagnostics is well calibrated for 
log O/H $>$ $-$3.5. At higher metallicities, the calibration is largely
derived from photoionization models.
\item The accuracy of each of these calibrations is quite limited. For
$R_{23}$ the usual quoted uncertainty is $\pm$0.2 dex, which is roughly
the scatter; in the turnaround region, the uncertainty is significantly
larger. The accuracy of $S_{23}$ is probably about the same; although
there are few data points to pin down the scatter at the present time.
The scatter in [N~II]/H$\alpha$ is significantly larger, about $\pm$0.3
dex; most of this scatter is real, not observational. 
\item The strong-line abundance relations are subject to systematic
errors, because the forbidden-line strengths depend on the stellar
effective temperature and ionization parameter as well as abundances.
If a galaxy has a low star formation rate and only low-luminosity
H~II regions with cooler O stars, the empirical calibration could give 
a systematically different O/H than a galaxy with many of the most
massive O stars and luminous giant H~II regions. 
\end{enumerate}

\subsubsection{Photoionization Models}

Some have said that the use of photoionization models to estimate
nebular abundances is the ``last resort of scoundrels'', so to speak. 
I admit that I have used photoionization models to commit offenses 
in the past. Since Grazyna Stasi\'nska will cover the mechanics of
photoionization modeling in detail in her lectures, I will confine
my remarks to what I feel are major uncertainties and observational
considerations that need to be addressed.

{\it Gas-star geometry:} This is an observational consideration, since
the geometry influences the ionization parameter. The classical model
of an H~II region is a uniform sphere with a point source of ionizing
photons. High-resolution images of real H~II regions show that they
are anything but this. The Orion Nebula is better modeled as a blister,
with the brightest areas being the photoevaporating surface of molecular 
cloud. {\it Hubble Space Telescope} images of giant H~II regions show them
to resemble bubbles more than filled spheres, often with a surrounding
halo of superbubbles. Young star clusters can often be found outside
the main H~II region associated with the superbubbles (Hunter et al.
1996). In many cases the ionizing 
cluster is not centrally condensed, but rather quite loose and extended 
(for example, NGC 604 and I Zw 18). Even the 30 Doradus nebula, which
is dominated by the compact cluster R136a, also includes an extended 
distribution of O giants, supergiants and Wolf-Rayet stars, plus 
luminous embedded, possibly pre-main-sequence stars (see Walborn 1991,
Bosch et al. 1999). To my knowledge, there has been no investigation 
of the effect of an extended distribution of ionizing sources on H~II 
region spectra. 

The effects of density and density variations are related to this 
problem. $T_e$, and thus the optical forbidden line strengths, are
very sensitive to density because of collisional de-excitation of
the far-IR fine structure cooling lines (Oey \& Kennicutt 1993). 
This is most true for metal-rich H~II regions, so, for example the
relation between $R_{23}$ and O/H from ionization models depends on
the average density asssumed. It is not clear yet that observed
integrated densities from, say, the [S II] line ratio accurately
reflect mean densities. It is highly likely that a range of densities 
is more representative of nebular structure, although a functional
form is not yet known. 

A great deal of work needs to be done in this area.

{\it Wolf-Rayet stars:} There are a number of myths about Wolf-Rayet
stars and their influence on the H~II regions. The first myth is that
the presence of Wolf-Rayet stars indicates an age of at least 3 Myr
for the ionizing OB association, which is the result obtained from
stellar evolution and spectrum synthesis modeling. However, this idea
is demonstrated to be not true specifically in the case of 30 Doradus, 
which contains numerous W-R stars, yet has a color-magnitude age of 
1.9 Myr (Hunter et al. 1995). This contradiction is the result of a
new population of hydrogen-rich W-R stars noted by de Koter, Heap \&
Hubeny (1997). Thus we need to reconsider our ideas about using W-R 
stars to constrain the ages of stellar populations.

\begin{figure}[t!]\label{fig_dustabs}
\resizebox{\hsize}{!}{\includegraphics{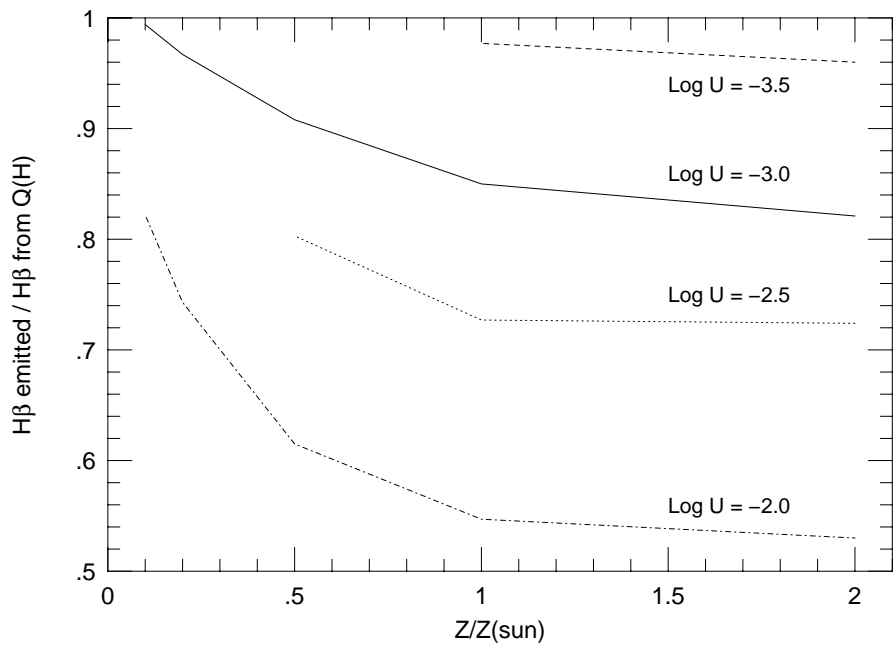}} \hfill
\parbox[b]{\hsize}{
\caption{The ratio of emitted H$\beta$ emission to that predicted from the
stellar ionizing photon luminosity as a function of the nebular abundances, 
showing the effects of absorption of ionizing photons by dust grains. The
models are for a stellar temperature of 40,000 K and assume a linear increase 
of the dust-to-gas ratio with metallicity. From Garnett (1999)}} 
\end{figure}

The second myth is that W-R stars add a hard component of photons
with energies greater than 54 eV to the ionizing radiation field. 
Again, this is a result from the combination of stellar atmosphere
models for Wolf-Rayet stars and spectrum synthesis models. The 
reality is that only about 1 in 100 W-R stars emits significant 
amounts of radiation beyond 54 eV. Those W-R stars that are associated
with nebular He~II emission in nearby galaxies tend to be rare high
excitation WN and WO types (Garnett et al. 1991). X-ray binaries are
also implicated in nebular He~II emission (Pakull \& Angebault 1986).
We do not yet understand the evolutionary status of these stars, so 
it is premature to predict them from the stellar evolution models.
Indeed, a comparison of photoionization models with the spectral
sequence of H~II regions indicates that OB cluster models that include 
such hot W-R stars produce results that are not consistent with 
observed emission-line trends (Bresolin, Kennicutt, \& Garnett 1999).

It is not clear that we know very well at all the ionizing spectral
energy distribution of Wolf-Rayet stars, or of O stars for that matter.
Stellar atmosphere models which incorporate stellar winds, departures
from LTE and plane-parallel geometry, and realistic opacities are 
still in the development stage, and the effective temperature scale
of O stars is still in flux (Martins, Schaerer, \& Hillier 2002). 

A great deal of work needs to be done in this area.

{\it Dust:} 
Dust has three major effects on the H~II region spectrum. First, dust 
grains mixed with the ionized gas absorb Lyman-continuum radiation. 
Second, obscuration by dust is typically patchy; differential extinction 
between stars and gas can affect the emission line equivalent widths.
Third, dust can affect the heating and cooling by emitting and recombining
with photoelectrons. 

The absorption cross-section for standard interstellar dust grains extends
well into the EUV spectral region with a peak near 17 eV. Dust grains are 
thus quite capable of absorbing ionizing photons in the H~II regions, and
in fact can compete with H and He. When this occurs, the flux of Balmer
line emission is reduced over the dust-free case. Figure 3 displays a set 
of ionization models showing the reduction in H$\beta$ line emission over
that expected from the number of ionizing photons for dusty H~II regions.
I have assumed standard interstellar grains (Martin \& Rouleau 1990), with
a dust-to-gas ratio that varies linearly with metallicity over the range
0.1-2.0 solar O/H. The models show that grains can reduce the emitted H$\beta$
flux by as much as 50\%. The amount lost depends strongly on the ionization
parameter, increasing for higher ionization parameters. A region with high
$U$ is likely to be a young one where the gas is close to the star cluster;
thus more H$\beta$ photons are missed, and EW(H$\beta$) reduced the most,
for the youngest clusters. 

Incidentally, the same phenomenon leads one to underestimate the number
of ionizing photons. Therefore, claims of leakage of ionizing photons
from H~II regions, based on comparing N(Ly-c) from Balmer lines fluxs 
with that estimated from the OB star population, must be viewed with
some skepticism.

\begin{figure}[t!]\label{fig_dustlines}
\resizebox{\hsize}{!}{\includegraphics{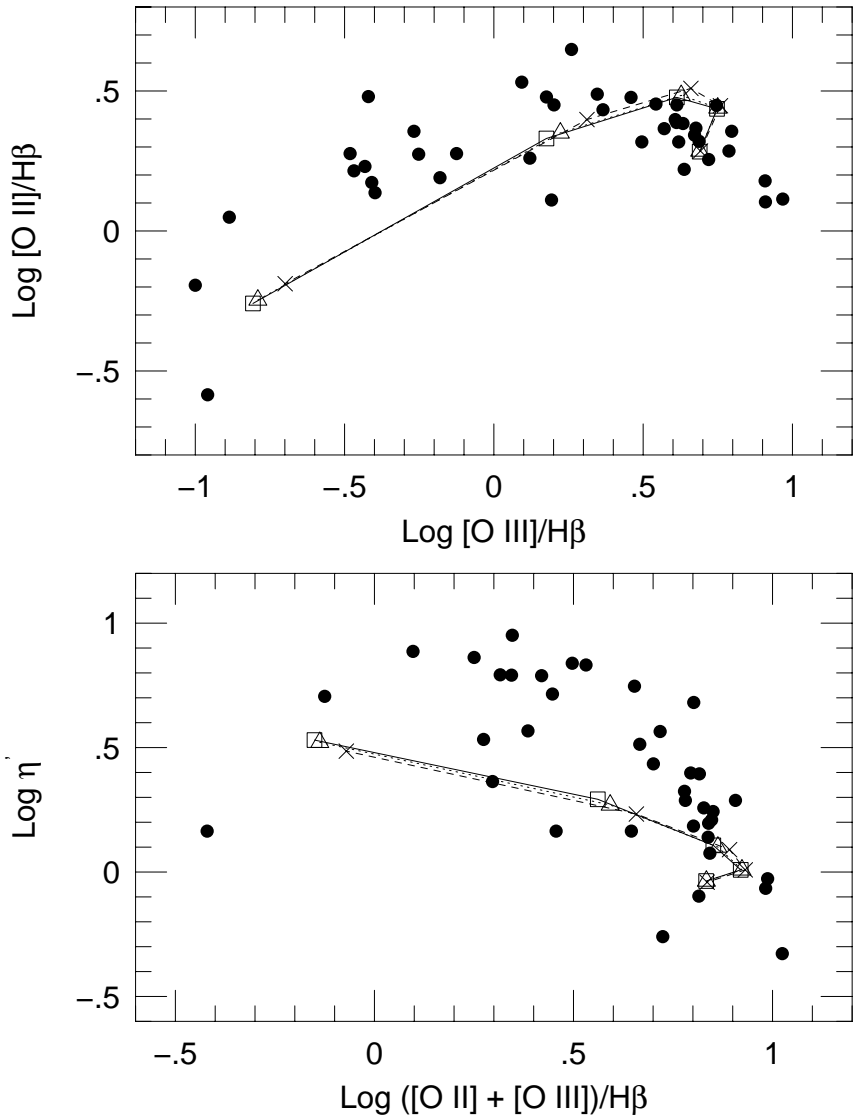}} \hfill
\parbox[b]{\hsize}{
\caption{The effects of dust grains on forbidden-line strengths in H~II
regions. Three sequences of models are shown, with $T_{eff}$ = 40,000 K
and log $U$ = $-$3. {\it Solid line plus squares:} dust-free models; 
{\it dashed line + crosses:} models with standard ISM grains; {\it dotted 
line plus triangles:} models with Orion-type grains. The effects of grains 
on the emission-line ratios are seen to be modest. From Garnett (1999)}}
\end{figure}

Differential extinction between the stars and the gas can also affect 
EW(H$\beta$).
Calzetti et al. (1994) found that the obscuration toward starburst clusters
tended to be lower than that toward the ionized gas. They determined that,
on average, $A_V$ toward the stars was about one-half of that toward the
gas. This is understandable if the stars have evacuated a cavity in the
ionized gas through the combined effects of radiation pressure and stellar
winds. The average derived obscuration for H~II regions in spirals is $A_V$
$\approx$ 1 mag. If $A_V$(stars) is only 0.5 mag, then the observed EW(H$\beta$)
will be about 40\% lower than the intrinsic value. 

These results suggest that dust effects can easily cause one to underestimate
the intrinsic EW(H$\beta$), even for metallicities as low as 0.1 solar O/H.
This would lead to a systematic bias toward larger ages for the stellar 
population. One should therefore exercise caution in weighting EW(H$\beta$) 
as a constraint on the synthesis models.

By contrast, the effects of dust on the relative forbidden-line strengths are
modest (Figure 4), except at high metallicities (Shields \& Kennicutt 1995). 
One exception is in the case of very hot ionizing stars (for example the
central stars of planetary nebulae), where ionization of grains can lead to 
additional photoelectric heating of the nebula (e.g., Ferland 1998, Stasi\'nska
\& Szczerba 2001). 

A great deal of work is needed in this area. 

{\it Note on atomic data for ionization calculations: the vast majority
of the values used for photoionization and recombination cross-sections 
are computed, not experimental.} This does not mean that their uncertainties 
are zero! In fact, the best values are probably not accurate to better than
15-20\%. Thus, it is unreasonable to expect photoionization models to 
match real H~II region spectra to an accuracy much better than this.

\subsection{Spectroscopy of Individual Stars}

Stellar spectroscopy has been a very valuable tool for studying the
composition and evolution of stars in our Galaxy. Recent improvements
in instrumentation and the construction of 8-10m telescopes has allowed
this kind of work to be extended to other galaxies. It is not possible
yet to do routine spectroscopy of F and G main sequence stars outside
the Milky Way, so these studies have concentrated on A and B type 
supergiants or red giants. Nevertheless, detailed abundance studies 
of individual stars is not likely to extend far beyond the Local Group
for some time because of telescope size limitations.

The supergiants are an important complement to spectroscopy of H~II
regions, since they sample similar spatial and temporal distributions.
Furthermore, they overlap in many of the elements that can be studied:
C, N, O, Ne, and so on. On the one hand, the supergiants provide 
information on elements such as Si, Fe and s-process elements that 
are depleted into grains in the ISM. On the other hand, the H~II regions 
provide a valuable comparison for He, C,  and N which may be affected 
by internal mixing and nucleosynthesis in the massive stars, which is 
covered by Norbert Langer's contribution. This is one of the important 
uncertainties in abundance studies for these stars; others include the 
degree to which conditions depart from LTE, and the effects of spherical 
geometry and stellar winds.

Spectroscopy of red giants is well established from Milky Way studies,
as discussed by David Lambert. Red giants are valuable because they
sample abundances over long time spans, from a 100 Myr to greater than
10 Gyr. A wide variety of elements can be studied, including $\alpha$
capture elements, Fe-peak elements, and neutron-capture elements. 
C, N, and O (and s-process elements in AGB stars) can be affected by 
internal mixing. Since red giants are fainter than H~II regions or 
supergiants, detailed spectroscopy is limited to the Milky Way's 
satellite galaxies at present.  

For metallicity distributions, one can examine lower spectral resolution
diagnostics. The most useful of these has been the Ca II triplet
indicator (Armandroff \& Da Costa 1991), which uses the combined
equivalent width of the Ca II triplet near 8500 \AA, calibrated with
metallicities of globular clusters, to infer the metallicity [Fe/H]
(where the brackets denote the logarithmic abundance relative to that
in the Sun).
The main uncertainty of this method is that the Ca/Fe abundance ratio
can vary depending on the star formation history, so the globular 
clusters may not provide the correct metallicity calibration for 
galaxies with a variety of star formation histories. Work needs to 
be done to calibrate the Ca II triplet with [Ca/H] rather than [Fe/H] 
to remove this ambiguity.

\subsection{Stellar Photometry and Color-Magnitude Diagrams}

Color-magnitude diagrams of galaxy populations can provide some
information on the metallicity (or metallicity spread) of a stellar
population, since features in the CMD, such as the color of the
red giant branch, can vary with metallicity. Unfortunately, these
features also vary with age of the populations, so there is a 
degeneracy between age and metallicity in the CMD (and in composite
colors). Systematic uncertainties may also be introduced by variations
in element abundance ratios and by reddening. Thus, color-magnitude
diagrams are at best indicative of metallicities.

\subsection{Spectrum Synthesis of Stellar Populations}

Spectrum synthesis for deriving metallicities has been applied 
mostly to elliptical galaxies. Since I am not an expert on this,
I refer the reader to Guy Worthey's review in Henry \& Worthey (1999)
and references therein for details. Needless to say, spectrum 
synthesis is an intricate and uncertain art; the results depend
on the choices of spectral templates, element abundance ratios, 
and star formation histories. Worthey points out that 25 spectral
indices are available to derive metallicities and ages for old
stellar populations. (In 1986 there were 11 indices in the Lick
spectral index system [Burstein, Faber, \& Gonz\'alez 1986].) 
Unfortunately, most of these vary in a degenerate way with age 
and metallicity of the stellar population. The best indices for
breaking this degeneracy are (1) H$\beta$ or a higher Balmer line,
which are more sensitive to age; and (2) Fe4668 (which is actually
a C$_2$ feature), more sensitive to metallicity. In addition, the
Mg index (covering Mg b and Mg$_2$ features between 5150 and 5200 \AA)
provides information on Mg/Fe. 

\subsection{Surface Photometry and Galaxy Colors}

\begin{figure}[t!]\label{fig_bellbrbv}
\resizebox{\hsize}{!}{\includegraphics{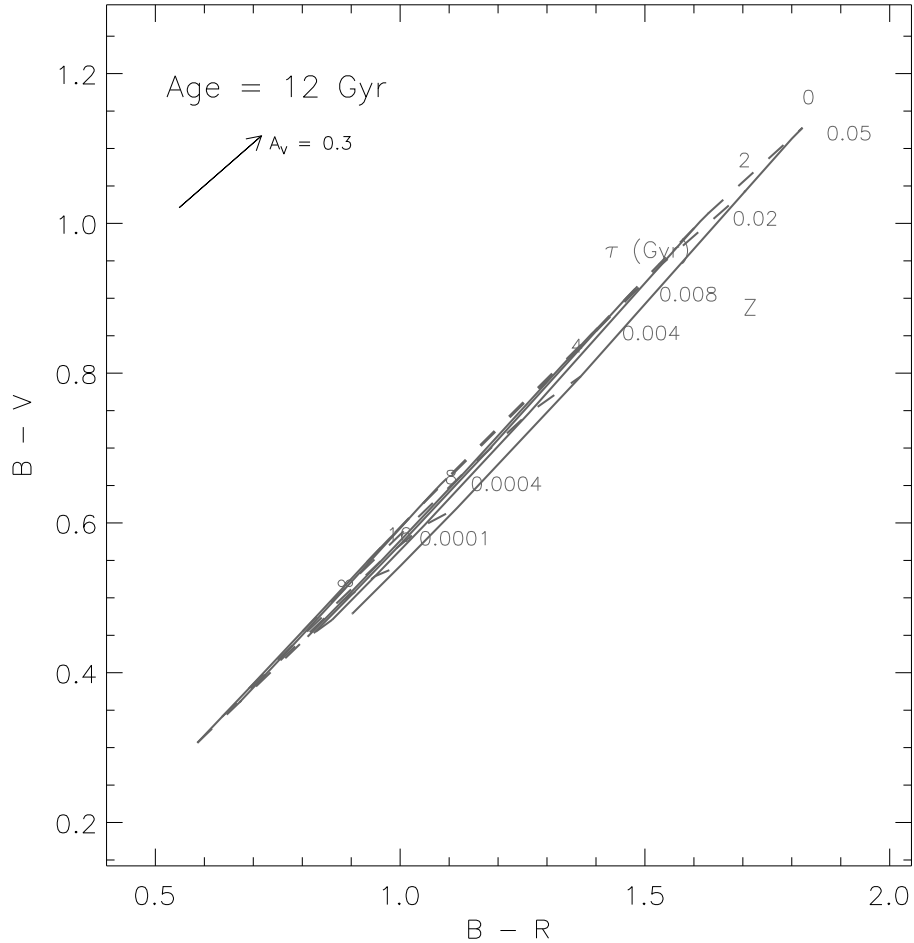}} \hfill
\parbox[b]{\hsize}{
\caption{Synthetic B--R vs. B--V colors for disk galaxies based on population 
synthesis models. The colors are from Bruzual \& Charlot models with an 
exponentially decaying star formation rate with timescale $\tau$ ranging
from 0 Gyr to infinity and metallicity Z ranging from 0.0001 to 0.05; see 
Bell \& de Jong (2000) for details of the models. Solid lines connect 
models of constant $Z$, while dashed lines connect models of constant 
$\tau$. All galaxies start forming stars 12 Gyr ago. The effects of 
interstellar reddening by foreground screen with $A_V$ = 0.3 magnitude 
is shown by the vector in the upper left corner. Diagram courtesy of 
Eric Bell. 
}}
\end{figure}

Since the colors of stars are sensitive to both age and metallicity
(as pointed out in Section 2.3 above) it is readily concluded that
the colors of galaxies similarly depend on age and metallicity. The 
situation is a bit more complicated because in galaxies the colors 
represent composite stellar populations. A further complication is 
the ubiquitous presence of dust, which has a clumpy distribution
mixed with the stars rather than a uniform screen. Nevertheless,
the analysis of galaxy colors could provide a useful means of 
studying averaged ages and metallicities for stellar populations
in very large samples of galaxies covering a wide range of redshifts,
particularly galaxies that are too faint for spectroscopy.

\begin{figure}[t!]\label{fig_bellmagbin}
\resizebox{\hsize}{!}{\includegraphics{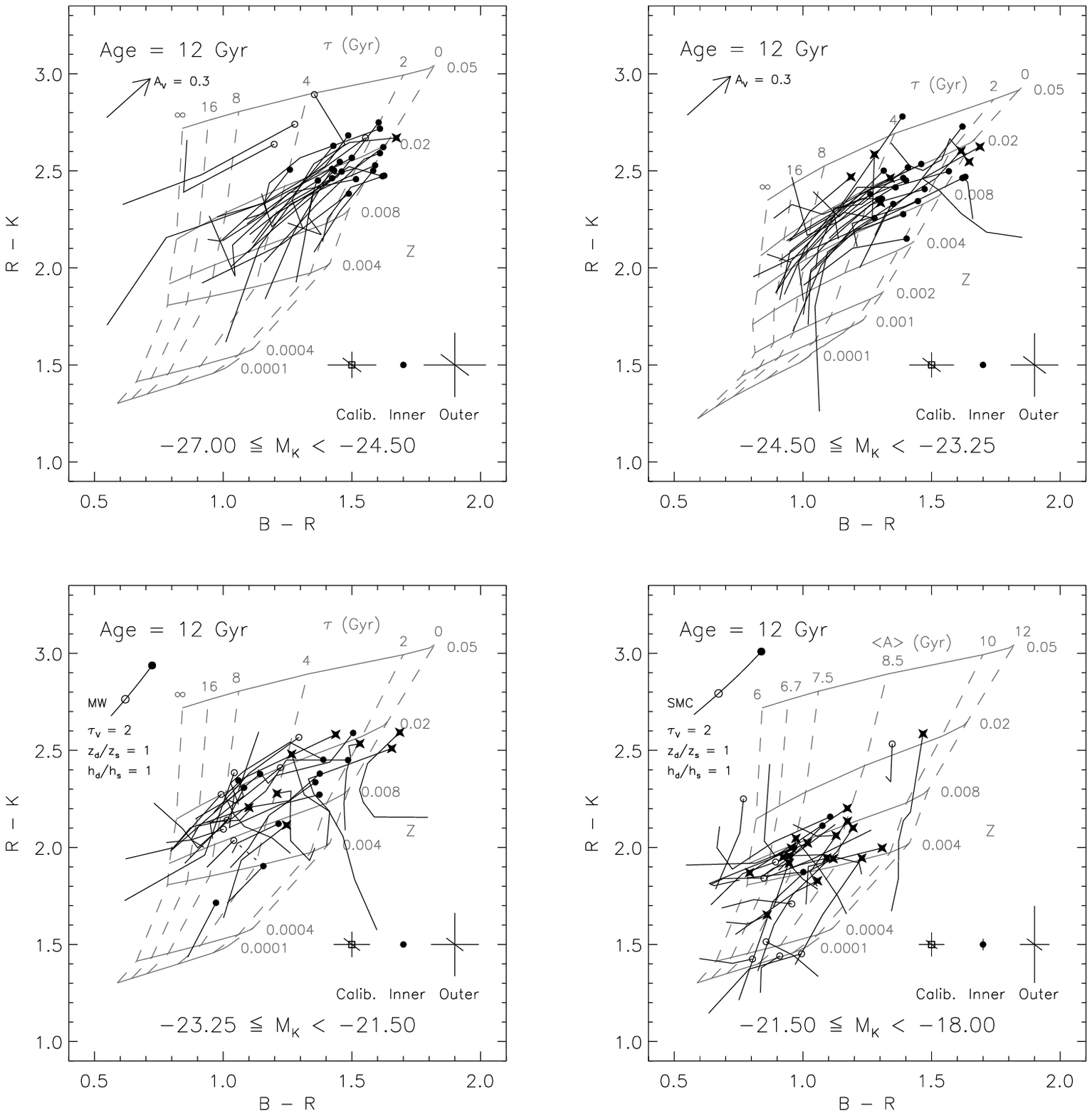}} \hfill
\parbox[b]{\hsize}{
\caption{Synthetic B--R vs. R--K colors for disk galaxies from Bell \& 
de Jong (2000); see this paper for complete details. Here the four
panels show results for galaxies in different ranges of K-band absolute
magnitude. The labels on the model grid are the same as in Figure 5; 
note that the bottom right panel shows the mean age $<A>$ rather than 
the star formation timescale $\tau$. The solid broken lines in each 
plot represent observed surface colors for disk galaxies in the Bell
\& de Jong sample; the attached open or filled circles are the central
colors for each galaxy. The symbols in the bottom right of each panel
show the calibration and sky subtraction error bars for the inner and
outer annuli of a galaxy. Diagram courtesy of Eric Bell. 
}}
\end{figure}

As with stars, of course, a difficulty with using colors is that
variations in age and metallicity cause similar variations in galaxy 
colors. This is especially true for optical colors, which have been
known for some time to be almost completely degenerate with regard 
to variations in age and metallicity (see Figure 5). This degeneracy
can be broken to a large extent by including IR photometry, particularly
K-band surface photometry, as illustrated in Figure 6 from Bell \& de 
Jong (2000). Bell \& de Jong have exploited this property to derive 
{\it luminosity-weighted} mean ages and metallicities for a sample of 
low-inclination disk galaxies. Note that the luminosity weighting
means that the derived properties do not represent the {\it present-day}
metallicities, as the emission-line measurements do. 

Note that because this method depends on synthesis models for the colors
of the stellar population, it suffers the same limitations. The model
colors depend on the star formation and chemical enrichment history of
a given galaxy. At present very simple star formation histories are 
assumed: either an instantaneous burst or exponentially decaying
continuous star formation (which approximates a constant star formation
rate for very long decay timescales). These approximations may break
down in cases of galaxies which have undergone multiple starbursts 
separated by long periods, or galaxies which have truncated star 
formation histories, possibly punctuated by starbursts as well. Dwarf
galaxies, in particular, may not be well-reproduced by the synthesis
models. Note also that the sensitivity of the color-color diagram 
decreases rapidly for very metal-poor or very old stellar populations.

Note also that the mean ages and metallicities of a given position are 
very sensitive to the dust correction, although the slopes of age and 
metallicity gradients are not. Finally, the uncertainties in photometry
and sky subtraction grow as the surface brightness decreases, so inferred 
ages and metallicities become increasingly more uncertain in the outer
parts of disks and in low surface brightness galaxies. In contrast,
metallicity gradients derived from H~II regions are more stable, as 
the luminosity of an H~II region is largely independent of its position 
within a galaxy. 

\section{Abundances in Local Group Dwarf Elliptical Galaxies}

Outside of the Magellanic Clouds, dwarf elliptical galaxies (dEs, 
sometimes also called dwarf spheroidals, or dSph) are the nearest 
companion galaxies to the Milky Way. The discovery of the tidally 
distorted Sagittarius dwarf elliptical (Ibata et al. 1994) brought
the number of known dE companions for the Milky Way to nine. There 
may be more still hidden behind the Galaxy's obscuring dust layer.
Recent studies have uncovered a host of dE companions of our sister
galaxy M31 as well, and a few other more distant 'free-floating' dEs, 
such as the Cetus galaxy (Whiting et al. 1999) have been found on 
Schmidt survey plates. Many of the properties of the Local Group
dEs have been tabulated by Mateo (1998). 

The dEs are deceptively simple stellar systems, with no young stars
and apparently kinematically-relaxed stellar populations. Recent
high-precision ground-based and HST CCD photometry have demonstrated
that this is far from the truth. The dEs display a variety of complex
multi-episode star formation histories. For example, Carina shows
evidence for several distinct star formation events spread over 
several Gyr (Smecker-Hane et al. 1994); Sculptor and Fornax appear 
morphologically similar, but Sculptor appears to formed the bulk of 
its stars at an earlier time than Fornax (Tolstoy et al. 2001). Only 
Ursa Minor appears to have something like a simple, monometallic 
stellar populations based on its color-magnitude diagram (Mighell \&
Burke 1999, but see below).

These star formation histories are of vital interest to understanding
the evolution of the dEs, and they have raised some puzzles. Among
these are the question of how, on the one hand, the dEs lost their gas, 
and how, on the other hand, they retained gas to experience multiple 
episodes of star formation! Combining the star formation history
with the element enrichment history can, in principle, yield the
information needed to understand the evolution of dwarf galaxies
and their contribution to enrichment of the IGM. 

\subsection{Metallicities}

In the absence of spectroscopy, metallicities for dEs have been derived 
from their CMDs by comparing the colors of the red giant branches (RGB) 
with those of globular clusters. This is not entirely satisfactory because 
of the age-metallicity degeneracy in RGB colors, and because element
abundance $ratios$, which also affect RGB colors, may not be the same
in the dEs as in the globulars. A few efforts have been able to estimate
metallicities from low-resolution spectra obtained with 4-5 meter 
telescopes; these studies are summarized in Mateo (1998). The photometric
and spectroscopic studies show the dEs to have quite low $<$[Fe/H]$>$,
ranging from about --1.3 in Fornax to about --2.2 in Ursa Minor. Most
of the dEs show evidence for a significant spread in [Fe/H], 
$\sigma$([Fe/H]) = 0.2-0.5 dex, based on the width in color of the RGB
(disregarding possible age spread contributing to this).

\begin{figure}[t!]\label{fig_sclsfh}
\resizebox{\hsize}{!}{\includegraphics{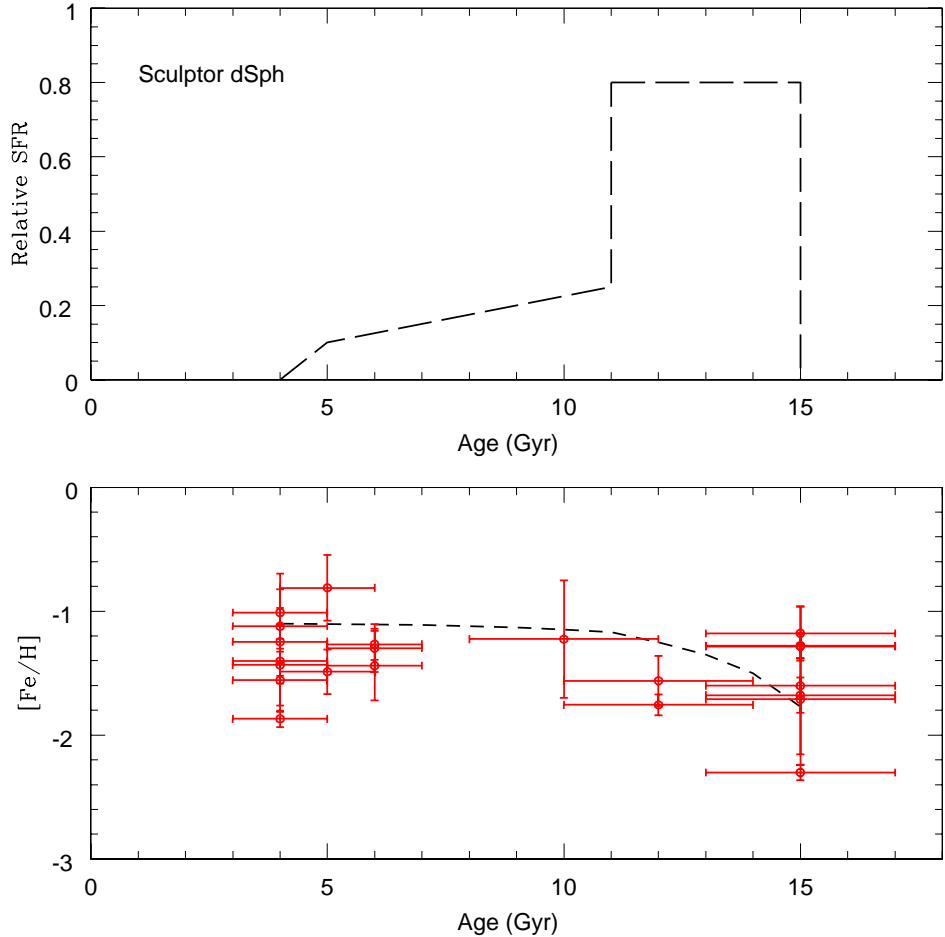}} \hfill
\parbox[b]{\hsize}{
\caption{
The star-formation and metallicity evolution history for Sculptor 
derived by Tolstoy et al. (2001) from Ca II triplet measurements
and photometry of red giants. The upper panel shows a schematic plot 
of how the star formation rate may have varied over time. The lower 
panel shows the corresponding variation in metallicity over the same 
time period ({\it dashed line}). Overploted on the lower panel are 
the Ca~II triplet measurements for individual Sculptor giants, with 
ages determined using isochrones.
}}
\end{figure}

With several 8-10 meter telescopes now available, medium-resolution 
(R $\approx$ 5,000) spectroscopy of RGs in the Milky Way satellite dEs 
can be almost routinely done, while high-resolution spectroscopy (R $>$ 
15,000) is possible for the brightest giants. These facilities offer 
exciting new possibilities for understanding the evolution of the Local 
Group dEs. One example of what can be done is the VLT study of Tolstoy
et al. (2001), who obtained Ca II triplet measurements for 37 red giants
in Sculptor, 32 RGs in Fornax, and 23 RGs in the dI NGC 6822. Having
measured metallicities for individual stars and existing CMDs for these 
galaxies, it was possible for Tolstoy et al. to assign ages to each star 
directly by comparison with isochrones of the proper metallicity, and 
subsequently to derive the time evolution of both the star formation 
rate and the metallicity. This is shown for Sculptor in Figure 7 taken 
from Tolstoy et al. The results are consistent with an initial burst of 
star formation between 11 and 15 Gyr ago with a metallicity [Fe/H] 
$\approx$ --1.8, a sharp subsequent decline in the SFR, followed by a 
slow decline in star formation until it stops approximately 5 Gyr ago. 
The metallicity evolution is very modest, with the youngest stars having 
a mean [Fe/H] of only --1.4. Their data for Fornax, on the other hand, 
show a low star formation rate over the time period 10-15 Gyr ago, then 
a sharply increased rate over the next few Gyr, and a higher mean [Fe/H] 
of --0.7 for the youngest stars. 

The main source of uncertainty here is that the Ca/Fe ratio in these
stars may not reflect that of the calibrator globular clusters stars -
and the Ca/Fe ratio may actually vary with time within each galaxy!
Nevertheless, with more data like this for the Local Group dwarfs 
it should be possible to derive very accurate evolution histories
for these galaxies.

\subsection{Element Ratios}

Element abundance ratios are another important piece of information,
since the various elements are synthesized in stars with different
masses and lifetimes. The $\alpha$-element/Fe ratio, in particular
is a good diagnostic, because the $\alpha$-capture elements are 
produced mainly in very massive stars and expelled into the ISM by
Type II or Ib,c supernovae, while Fe is mainly produced in Type I
supernovae by longer-lived stars. Thus the $\alpha$/Fe ratio is an 
indicator of how rapidly star formation and metal enrichment occurred 
within a system: high $\alpha$/Fe indicates enrichment over short 
time scales, possibly in starburst events, while low $\alpha$/Fe may 
indicate enrichment over much longer time scales, as in systems with 
roughly constant star formation rate over their lifetimes. 

Of current interest is the question of whether the Galactic halo was
formed in a monolithic collapse (Eggen, Lynden-Bell \& Sandage 1962) or 
was aggregated from mergers of smaller sub-units (Searle \& Zinn 1978). 
It is speculated that the nearby dE satellites may be representative
of those sub-units.

It is clear from their CMDs that many of the Milky Way satellites are 
not like the halo population, since they contain stars that are much 
young than those in the halo. At the same time, we know that the 
Sagittarius dE is being tidally disintegrated by the Galaxy and will 
become part of the halo. Although systems like Fornax and Carina 
appear not to be representative of the current halo populations, much 
less complex systems like Ursa Minor and Draco could be similar to 
the structures out of which the halo formed. 

The test of this possibility is that the ages and compositions of
stars in the dEs are similar to those in the halo. It is known that
halo stars show elevated [$\alpha$/Fe] compared to disk stars, 
reflecting a dominant nucleosynthesis contribution from massive 
stars, while [Ba/Eu] is low in halo stars, indicating that the 
$s$-process (which is the main source of Ba) has not had sufficient
time to contribute to the abundances in halo stars, while the 
$r$-process (the main source of Eu) dominates in metal-poor stars. 
Indeed, very metal-poor halo giants show evidence for a purely 
$r$-process contribution to the abundances of heavy neutron capture
elements (Sneden et al. 2000).

Little high-resolution spectroscopy of giants in even the nearest dEs 
have been carried out because of the faintness of the stars (16th-20th 
magnitude). However, the first such studies have become available 
due to the availability of the 10-m Keck telescopes (Shetrone, Cot\'e,
\& Sargent 2001). Shetrone et al. have obtained high-resolution spectra 
for 5-6 stars in each of the Draco, Ursa Minor, and Sextans galaxies, 
deriving abundances for a variety of elements. The comparison of element
abundance ratios in these stars show a puzzling mixed bag of results. 
Although Shetrone et al. argue that [$\alpha$/Fe] is low in the dEs
compared to halo stars, closer examination shows that [Ca/Fe] and
[Ti/Fe] do appear to be lower, but [Mg/Fe] and [Si/Fe] appear to agree 
with halo ratios. Meanwhile, [Ba/Eu] values in the dEs appear to be in 
good agreement with those in halo stars. The comparison between dEs and 
halo stars seems to be inconclusive at the present time, perhaps not 
surprising given the small samples of stars at present. The samples of 
stars for each galaxy certainly need to be enlarged to determine the 
evolution of element ratios in these galaxies. Nevertheless, the 
Shetrone et al. (2001) study illustrates the power of high-resolution
spectroscopy with the new large telescopes. More work of this nature
is highly encouraged. 

\section{Abundance Profiles in Spirals and Irregulars}

Here I present an overview of the patterns of metallicity and element
abundance ratios observed in spiral and irregular galaxies. I will 
discuss the results for both types of galaxies rather than separately;
many aspects can be discussed for the combined groups, although there
are a number of differences that could constitute the topic of an entire 
conference alone.

The observational data to be discussed represent a highly selected
sample of abundances, gas masses, and stellar photometry from sources 
too numerous to mention here. (If you recognize your data in the 
following plots, feel free to take credit.) I will employ
abundances derived almost exclusively from H~II region spectra, since 
they contribute the largest set of abundance data for spirals and 
irregulars in the local universe. 


\subsection{Gas and Stellar Masses}

The ultimate goal of any chemical evolution model is to account for
the global and local metallicity within a galaxy, the gas and stellar 
mass distributions, and the stellar luminosity self-consistently. Thus, 
any discussion of abundances and chemical evolution should include a
few words about observational determinations of gas and stellar masses 
and mass surface densities. 

\subsubsection{Neutral and Molecular Gas}

Neutral gas is the largest component of gas in most spirals and 
irregulars, as determined from H~I 21-cm hyperfine line measurements. 
The 21-cm line has been well-mapped in many nearby galaxies. With
regard to determining gas fractions and surface densities, a few
points should be kept in mind:
\begin{enumerate}
\item {\it The size of the neutral gas disk is often much larger than 
the stellar disk in spirals and irregulars}, as much as 3-4 times the
photometric radius $R_{25}$ (e.g., Broeils \& van Woerden 1994). One 
must obtain or use maps which cover the full extent of the H~I disk,
which can mean observing over degree-size scales for the nearest
spirals. Determining the H~I extent may be difficult in complicated
interacting systems such as the M81 group (Yun et al. 1994).
\item {\it Fully sampled maps are desirable}. Aperture synthesis
maps, while providing the high spatial resolution needed for studies
of kinematics and gas structure, can miss a large fraction of the
H~I emission on scales larger than the synthesized beam due to lack 
of short antenna baselines in the $u-v$ plane. (The closest spacings 
in an interferometer are, of course, one antenna diameter.) This 
is especially true for the highest resolution images. Although there 
are efforts to correct for the missing extended emission by including 
single-dish measurements, it is usually assumed that the extended 
emission is uniformly distributed. This assumption may not be correct.
\item {\it The helium contribution to the mass is not negligible.} 
The helium accounts for 30-40\% of the total gas mass, which must be 
included in the total gas mass and surface density. 
\end{enumerate}

Fortunately, many nearby galaxies have been well mapped in H~I at
kiloparsec scale resolution or better. 

Molecular gas, mostly in the form H$_2$, is the important phase associated 
with star formation. Although H$_2$ may not dominate the total gas mass,
it is often found to be the main component in the inner disk of Sbc or
later type spirals, and so can be the main contribution to the gas 
surface density in such regions. 

H$_2$ has no dipole moment and thus emits no strong dipole radiation of
its own. The usual tracer of molecular gas is the abundant CO molecule, 
typically the $^{12}$CO (J=1--0) millimeter-wave transition. The conversion 
from the measured I(CO) to the column density N(H$_2$),
X(CO), must be calibrated largely without the benefits of direct measurement 
of H$_2$ column densities. The result has been a long-standing controversy 
over the value of the CO-H$_2$ conversion factor and its dependence on
metallicity. For example, Wilson (1995) has studied the CO-N(H$_2$)
relation in a variety of environments in Local Group galaxies, comparing 
$I(CO)$ with molecular cloud masses derived from the velocity dispersion 
assuming the clouds are in virial equilibrium. From her data Wilson found 
a roughly linear relation between X(CO) and 12 + log O/H corresponding
to approximately a factor ten increase in X(CO) for factor ten decrease
in O/H, from solar O/H to 0.1 times solar O/H. On the other hand, 
Israel (1997a,b) argues that virial equilibrium is a poor assumption for
short-lived molecular clouds. He instead uses the FIR dust emission
surface brightness and H~I maps to determine the dust/N(H) ratio, then 
uses the FIR surface brightness to estimate N(H$_2$) in regions where
CO is detected. With this method Israel derives a variation in X(CO)
with O/H which is much steeper than that obtained by Wilson: a factor
of approximately 100 decrease in X for a factor 10 increase in O/H.

Both of these methods likely suffer from errors due to assumptions 
made. Virial equilibrium may very well be a poor assumption for 
molecular clouds. On the other hand, the FIR calibration depends on
the assumption that the dust-to-gas ratio is the same in neutral gas 
and in molecular clouds; however, grains may be preferentially 
destroyed by shocks in the lower density neutral component. The
FIR model is also very sensitive to the dust temperature. The relation
between I(CO) and N(H$_2$) likely depends on a variety of factors 
besides metallicity (Maloney \& Black 1988). More work needs to be 
done to determine the best method to obtain H$_2$ masses.

\subsubsection{Stellar Mass Densities}

Masses and surface densities for the stellar component in galaxies are 
probably even more uncertain than molecular gas masses. Because of the 
flat rotation curves of disk galaxies, and the consequent inference 
that the galaxies are dominated by non-luminous, non-baryonic matter, 
the mass-to-light (M/L) ratio and mass surface density of the stellar 
component can not be derived from galaxy dynamics. Unless one can count 
the stars in a region directly (possible only for very nearby systems), 
it is necessary to infer M/L for the stellar component by indirect means. 
This is difficult because the luminosity and colors of composite stellar 
populations depend on both the star formation history and the metal 
enrichment history. 

Nevertheless, recent work by Bell \& de Jong (2001) indicates that
the stellar M/L ratios of galaxies are rather robustly related to  
their colors. Bell \& de Jong examined population synthesis models
for galaxies assuming a variety of star formation histories. They
found that M/L for the stellar component correlated very well with 
optical colors, although there is some scatter in the correlations.
IR colors did not correlate as well with M/L because of the strong
metallicity dependence of the IR luminosity of giants. The B-band
M/L ratio shows a steep correlation with color, while the K-band 
M/L ratio shows a much less steep correlation (a factor three 
increase between B--R = 0.6 and B--R = 1.6, compared to a factor
10 increase in B-band M/L over the same color range). If the 
population synthesis models can reliably reproduce the colors
and spectra of real galaxies, this method offers the possibility
of greatly improved estimates of masses and mass surface densities 
for the stellar components of disk galaxies.

\subsection{Spatial Abundance Profiles}

\begin{figure}[t!]\label{fig_M101grad}
\resizebox{\hsize}{!}{\includegraphics{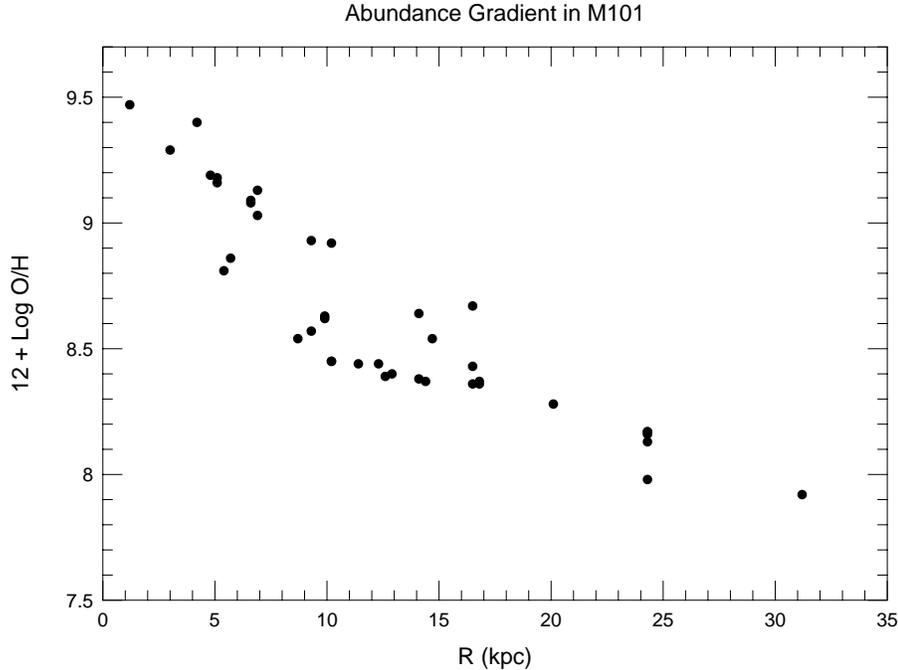}} \hfill
\parbox[b]{\hsize}{
\caption{The gradient in O/H across the disk of the spiral galaxy 
M101 vs. galactocentric radius (Kennicutt \& Garnett 1996). 
}}
\end{figure}

The spatial distribution of abundances in galaxies depends coarsely
on the Hubble type. Spectroscopic study of H~II regions in unbarred 
or weakly barred spiral galaxies typically reveals a strong radial 
gradient in metallicity (Figure 8), as determined from O/H (Zaritsky, 
Kennicutt, \& Huchra (1994; ZKH); Vila-Costas \& Edmunds 1992; VCE). 
The derived O/H can drop by a factor of ten to thirty or fifty from
the nucleus of a galaxy to the outer disk as demonstrated in galaxies 
with well-sampled data. In those spirals with spectroscopy of more 
than ten H~II regions covering the full radial extent of the disk, 
there is little evidence that O/H gradients deviate from exponential 
profiles. Irregular galaxies, by contrast, show little spatial 
variation in abundances, to high levels of precision (Kobulnicky 
\& Skillman 1996), indicating a well-mixed ISM. The data for strongly
barred spiral galaxies shows evidence that their O/H gradients are
more shallow than in unbarred spirals. I will discuss these galaxies
in more detailed in a later section.

A quick glance at data on O/H in galaxies (e.g. Figure 8 of ZKH) does 
not immediately reveal any trends of metallicity among galaxies of 
different types. However, detailed examination of this data shows that 
there are significant correlations between abundances and abundance 
gradients in spirals and irregulars with galaxy structural properties. 
Here I shall review some of these correlations and some implications. 
Note also that for the most part this discussion applies only to high 
surface brightness ``normal'' spirals.

\subsection{Metallicity versus Galaxy Luminosity/Mass}

\begin{figure}[t!]\label{fig_AReMV}
\resizebox{\hsize}{!}{\includegraphics{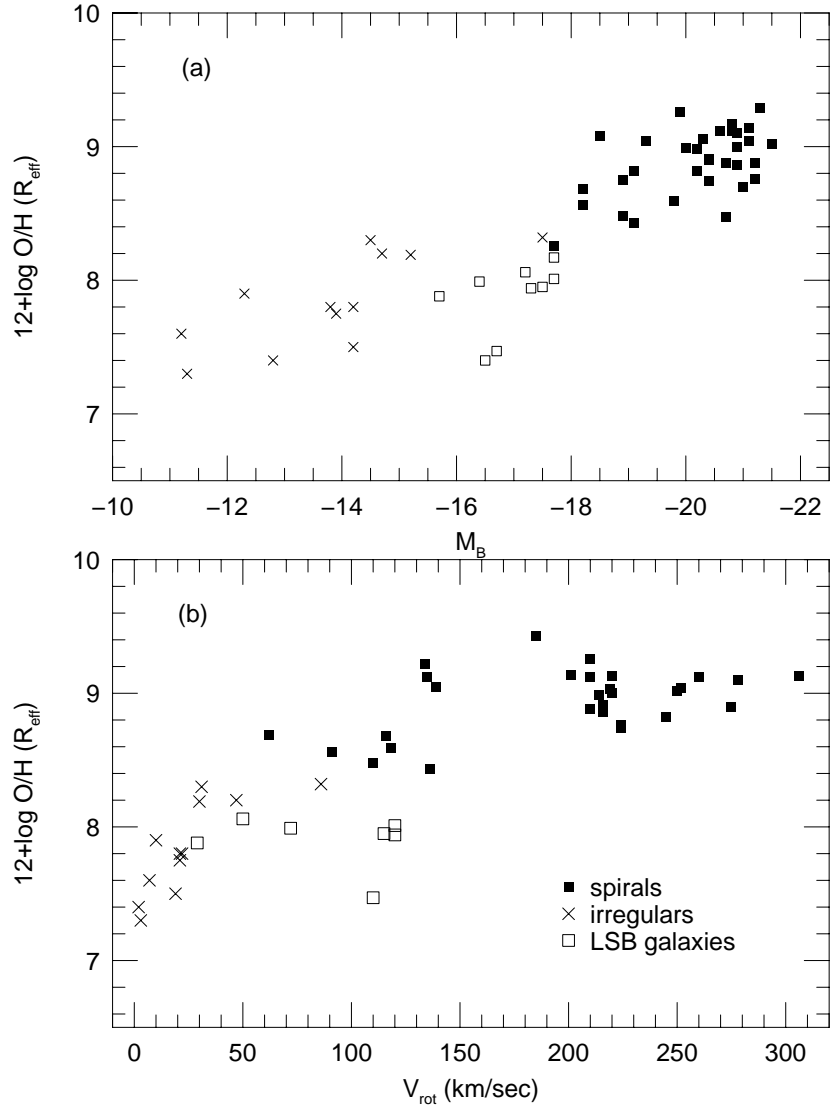}} \hfill
\parbox[b]{\hsize}{
\caption{Top: The correlation of spiral galaxy abundance (O/H) at the
half-light radius of disk from the galaxy nucleus vs. galaxy blue 
luminosity. Bottom: abundance versus maximum rotational speed $V_{rot}$. }}
\end{figure}

One well established correlation is the relation between metallicity 
and galaxy luminosity or (Garnett \& Shields 1987, Lequeux et al. 1979). 
This is shown in the top panel of Figure 9, where I plot O/H determined
at the half-light radius of the disc ($R_{eff}$)  versus B-band magnitude
$M_B$. This is the usual way of plotting the relation. [Note that the
choice of what value of metallicity to use for spiral disks, where the
metallicity is not constant, is somewhat arbitrary. I have used the
value at one disk scale length in the past on the grounds that the disk
scale length is a structural parameter determined by galaxy physics, whereas 
the photometric radius can be biased by observational considerations.
The actual ``mean'' abundance in the disk ISM would be determined by
convolving the abundance gradient with the gas distribution. As a simple
compromise I have used the disk half-light radius, which is 1.685 times
the disk scale length (de Vaucoleurs \& Pence 1978), and so is still
connected to galaxy structure. It should be noted also that a O/H - $M_B$
correlation is derived whether one uses the central abundance, the 
abundance at one disk scale length, or some other fractional radius.] 
ZKH noted the remarkable uniformity of this correlation over 11 magnitudes 
in galaxy luminosity, for ellipticals and star-forming spiral and irregular 
galaxies. 

To the extent that blue luminosity reflects the mass of a system, the
metallicity-luminosity correlation suggests a common mechanism regulating 
the global metallicity of galaxies. What the mechanism might be is not 
very well understood at present. The most commonly invoked mechanism is 
selective loss of heavy elements in galactic winds (e.g., Dekel \& Silk 
1986). However, the metallicity-luminosity correlation for star-forming
galaxies by itself does not imply that lower-luminosity galaxies are
losing metals. Such a correlation could occur if there is a systematic
variation in gas fraction across the luminosity sequence, either because
the bigger galaxies have evolved more rapidly, or because the smaller
galaxies are younger. In fact, there is evidence that both of these may 
be true. There is also evidence for fast outflows of hot X-ray gas from 
starburst galaxies such as M82 (Bregman, Schulman, \& Tomisaka 1995).
However, the question of whether this hot gas is escaping into the IGM
or will be retained by any given galaxy depends not just on the 
gravitational potential, but also on details such as the vertical 
distribution of ambient gas and the radiative cooling which are not
so well understood.

The question of loss of metals from galaxies is profound because of
the existence of metals in low column density Lyman-$\alpha$ forest 
systems (Ellison et al. 1999), which are probably gas clouds residing 
outside of galaxies. Where the heavy elements came from in these systems 
is still a mystery; it is possible that they were seeded with elements 
from a generation of pre-galactic stars or with elements expelled in 
starbursts during galaxy formation. In the dense environments of 
galaxy clusters, ram pressure stripping by intracluster gas, tidal 
interactions, and galaxy mergers may also liberate material to the 
intracluster medium (Mihos 2001). It is therefore a useful exercise to 
investigate what kinds of galaxies
are potential candidates for ejecting heavy elements into the IGM.

\begin{figure}[t!]\label{fig_YreV2}
\resizebox{\hsize}{!}{\includegraphics{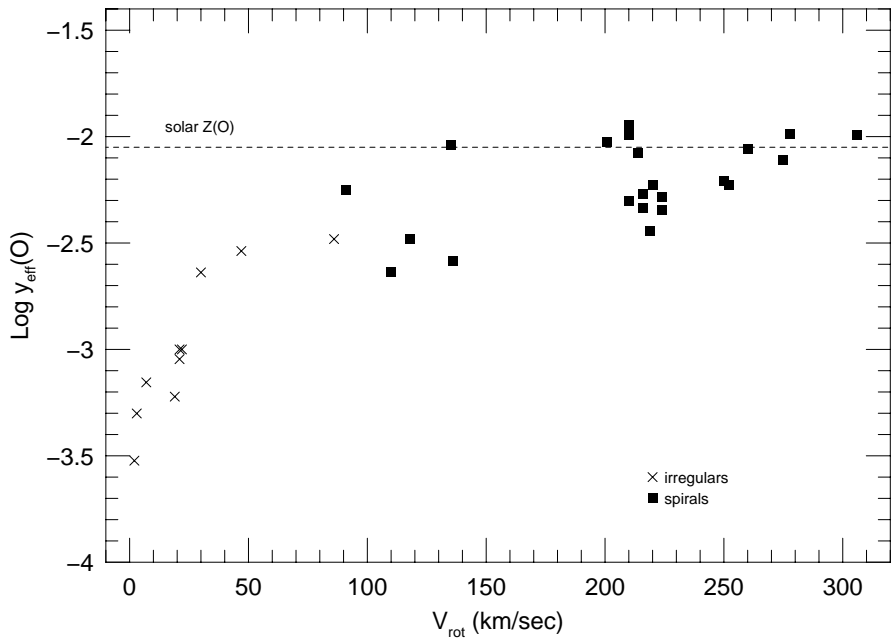}} \hfill
\parbox[b]{\hsize}{
\caption{Effective yields $y_{eff}$ for nearby spiral and irregular 
galaxies versus rotation speed $V_{rot}$ (Garnett 2002). Filled
squares represent the data for spirals while the crosses show 
the data for irregulars.}}
\end{figure}

I begin by looking at the metallicity-luminosity relation in a different
way. It is often argued that the B-band luminosity is not a very good
surrogate for mass, since the B-band light can be affected by recent
star formation and dust. Therefore, in Figure 9(b) I plot the mean
O/H for the galaxies in Fig. 9(a) versus rotation speed (obtained 
from resolved velocity maps [e.g., Casertano \& van Gorkum 1991, Broeils
1992], not single-dish line widths), where for the spirals the rotation 
speed is taken to be the value on the flat part of the rotation curve. 
Interestingly, the Z-$V_{rot}$ correlation turns over and flattens out
for $V_{rot}$ $\gtrsim$ 150 km s${-1}$, suggesting that spirals with
rotation speeds higher than this have essentially the same average
metallicity. Does this indicate a transition from galaxies that are
likely to be losing metals to the IGM to galaxies that essentially 
retain the metals they produce?

This question can be examined further by studying metallicity as a
function of gas fraction. In the context of the simple, closed box, 
chemical evolution, Edmunds (1990) derived a few simple theorems that 
show that outflows of gas and inflows of metal-poor gas cause galactic 
systems to deviate from the closed box model in similar ways. Specifically, 
defining the effective yield $y_{eff}$ 
\begin{equation}
y_{eff} = {Z\over ln(\mu^{-1})}, 
\end{equation}
outflows of any kind and inflows of metal-poor gas tend to make the 
effective yield smaller than the true yield of the closed box model. 
$y_{eff}$ defined this way is an observable quantity, and provides
a tool for studying gas flows in galaxies. Although the true yield
is relatively uncertain, comparing effective yields for a sample of
galaxies can provide information on the relative importance of gas
flows from one galaxy to another.

Such a comparison is presented in Figure 10 (Garnett 2002), where I 
show data compiled on abundances, atomic and molecular gas, and photometry
for 22 spiral and 10 nearby irregular galaxies. Figure 10 plots the 
effective yields derived for each
galaxy using equation 3.1 and the global gas mass fraction versus
galaxy rotation speed. The plot shows a very strong systematic 
variation in $y_{eff}$ with $V_{rot}$ in the dwarf irregular galaxies, 
with $y_{eff}$ increasing asymptotically to a roughly constant value
for the most massive galaxies. The uncertainties in the individual
$y_{eff}$ values are relatively large, because of relatively large
uncertainties in $M/L$ ratios for the stellar component and the 
CO - H$_2$ conversion for the molecular gas component; individual
values of $y_{eff}$ are probably not known to better than a factor 
of two. Nevertheless, the data show a factor of 30 systematic 
increase in $y_{eff}$ from the least massive irregulars to the most
massive spirals. 

This result is striking verification that the yields derived for dwarf 
irregulars are significantly lower than in spiral galaxies, and shows 
that the variation is a systematic function of the galaxy potential. In 
strict terms, the trend in Figure 10 does not distinguish between 
infall of unenriched gas and outflows as the cause. However, the trend 
toward small $y_{eff}$ in the least massive galaxies suggests that it is
the loss of metals in galactic winds that drives the correlation. It 
would be of interest to use this correlation to estimate the total amount
of gas lost in the small systems, and to determine the manner in which
supernova energy feeds back into the ISM of the host galaxies. 

Although not quite certain yet where one can say that galaxies are 
losing significant quantities of metals and which ones retain 
essentially all their metals, it appears likely that this boundary
point is somewhere near $V_{rot}$ $\approx$ 150 km s$^{-1}$. Given
this, one can surmise that the outflow of hot gas seen in the starburst 
galaxy M82 ($V_{rot}$ $\approx$ 100 km s$^{-1}$) may contribute 
significantly to enrichment of the IGM, while the hot gas flow seen
in NGC 253 ($V_{rot}$ $\approx$ 210 km s$^{-1}$) is likely to remain
confined to the galaxy.

\subsection{Abundance Gradient Variations }

\begin{figure}[t!]\label{fig_gradMB}
\resizebox{\hsize}{!}{\includegraphics{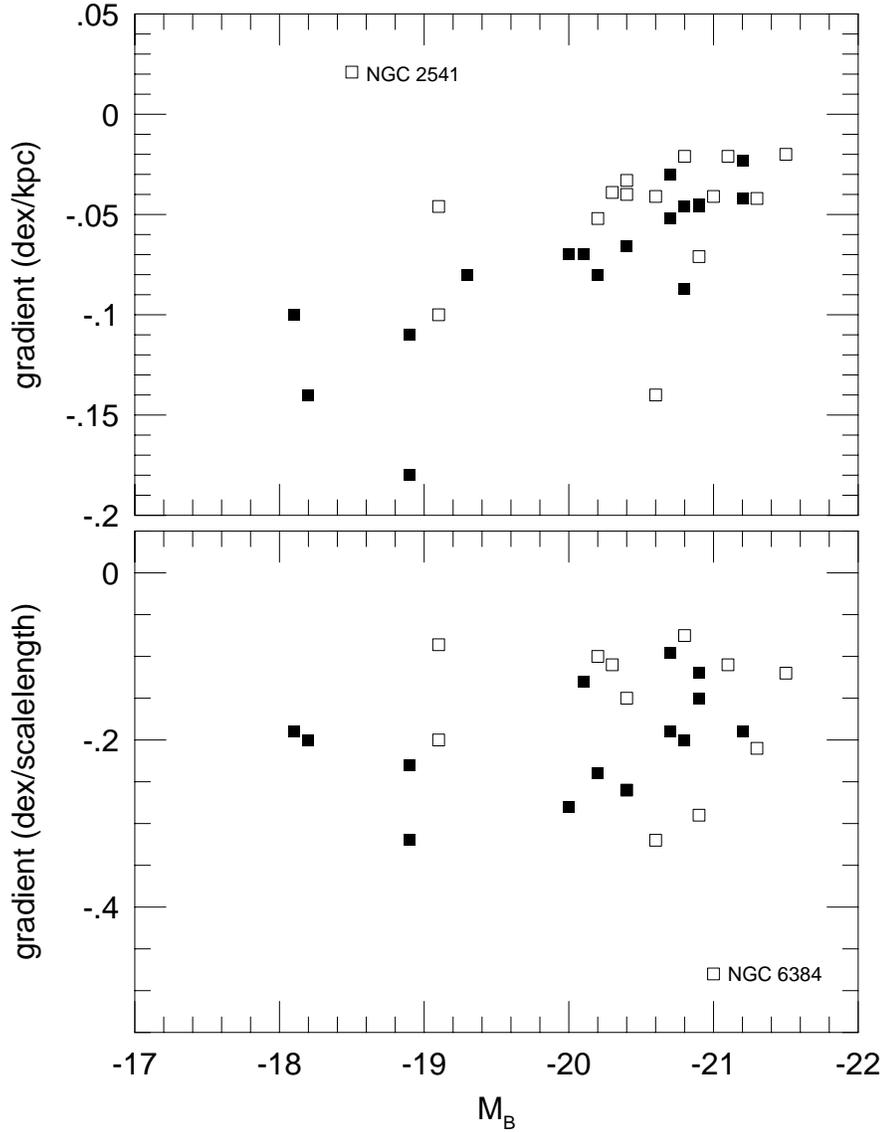}} \hfill
\parbox[b]{\hsize}{
\caption{ The correlation of abundance gradient vs. $M_B$, from Garnett et 
al. (1997a). The upper panel shows abundance gradients per kpc, while 
the lower panel shows gradients per unit disk scale length. }}
\end{figure}

Another commonly-noted correlation is illustrated in the top panel of Figure 
11: the steepness of abundance gradients (expressed in dex/kpc) decreases 
with galaxy luminosity. However, more luminous galaxies have larger disk 
scale lengths, and so if one looks at the gradient per disk scale length 
(Figure 11, bottom panel), the correlation goes away. Interestingly, when 
one considers the errors in the computed gradients (25\% is a typical 
uncertainty), then the scatter in measured gradient slopes may be consistent 
with purely observational scatter. Combes (1998) has suggested that a 
``universal'' gradient slope per unit disk scale length may be explained by 
so-called ``viscous disk'' models (Lin \& Pringle 1987); if the timescale 
for viscous transport of angular momentum is comparable to the star formation 
timescale (with the two timescales connected through the gravitational 
instability perhaps), such models naturally produce an exponential stellar 
disk. Chemical evolution models invoking viscous transfer have been able to 
produce abundance gradients (Clarke 1989; Tsujimoto et al. 1995), but it is 
not yet clear from the few models examined that they can yield very similar 
abundance gradients per unit disk scale length for a wide variety of spiral 
disks. This is something that deserves further study.

\begin{figure}[t!]\label{fig_prantzos1}
\resizebox{\hsize}{!}{\includegraphics{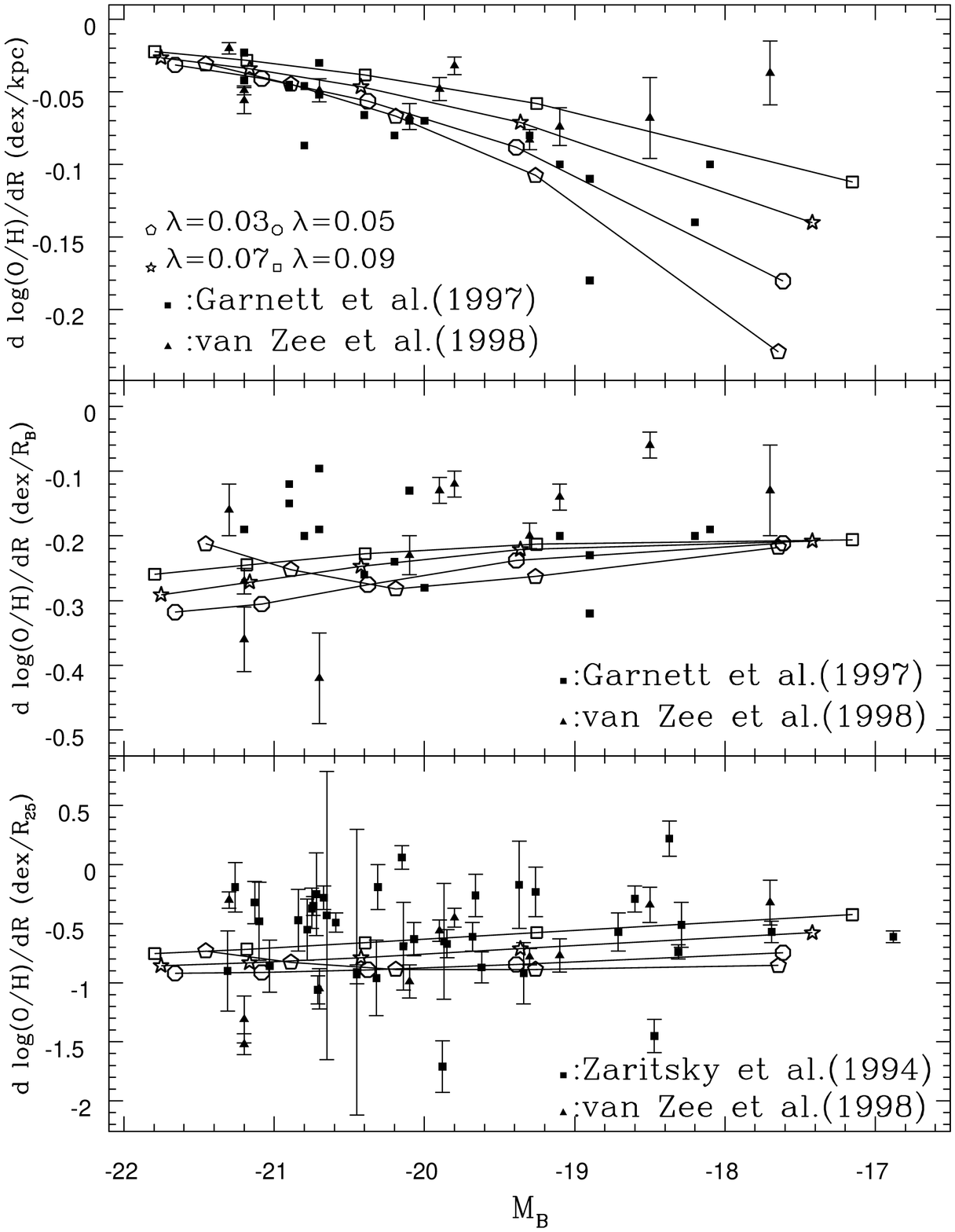}} \hfill
\parbox[b]{\hsize}{
\caption[]{Comparison of generalized chemical evolution models from 
Prantzos \& Boissier (2000) with observed composition gradients for
spiral galaxies. Top: O/H gradient in dex/kpc vs. $M_B$. Middle:
O/H gradient in dex per unit disk scale length vs. $M_B$. Bottom:
O/H gradient over the photometric radius $R_{25}$ vs. $M_B$. The
curves show the model relations for constant $\lambda$ for $\lambda$
= 0.03, 0.05, 0.07, and 0.09. 
}}
\end{figure}

Another interesting possibility is presented by Prantzos \& Boissier
(2000). They constructed a sequence of chemical evolution models for
disk galaxies by scaling the mass distribution (total mass, scale length)
according to the scaling relations for cold dark matter halos of Mo, Mao 
\& White (1998), in which the disk mass profile can be expressed using
only two parameters: the maximum circular velocity (which corresponds 
to the halo mass) and the spin parameter $\lambda$ (which corresponds
to the angular momentum). A key assumption is that the scaled galaxies
settle into exponential disks. Prantzos \& Boissier computed models for 
galaxies under these assumptions, calibrated by reproducing measurements
for the Milky Way. The basic results are illustrated in their Figure 4,
reproduced here in Figure 12. The top panel plots the slope of the
composition gradients in units of physical length (kiloparsecs) vs.
$M_B$, the middle panel plots the gradients per unit disk scale length,
while the bottom panel plots the gradient over the photometric radius
$R_{25}$. In each case the model gradients are in good agreement on
average with observed gradients, although the gradients per unit scale
length and $R_{25}$ are perhaps a bit steeper than observed. If this
analysis holds up it would be quite remarkable, as it would have been
difficult to imagine that the present-day ISM composition could be
related to the initial properties of the disks in the distant past. This 
may reflect the assumption that the baryons start out with exponential
mass distributions. Prantzos \& Boissier (2000) predict that there should 
be a small spread in observed gradients per kiloparsec in massive spirals 
and a large spread in small spirals, such that small spirals with large 
angular momentum should have shallower gradients (larger scale lengths) 
than those with lower angular momentum. 

\subsection{Metallicity vs. Surface Brightness}

\begin{figure}[t!]\label{fig_absurf}
\resizebox{\hsize}{!}{\includegraphics{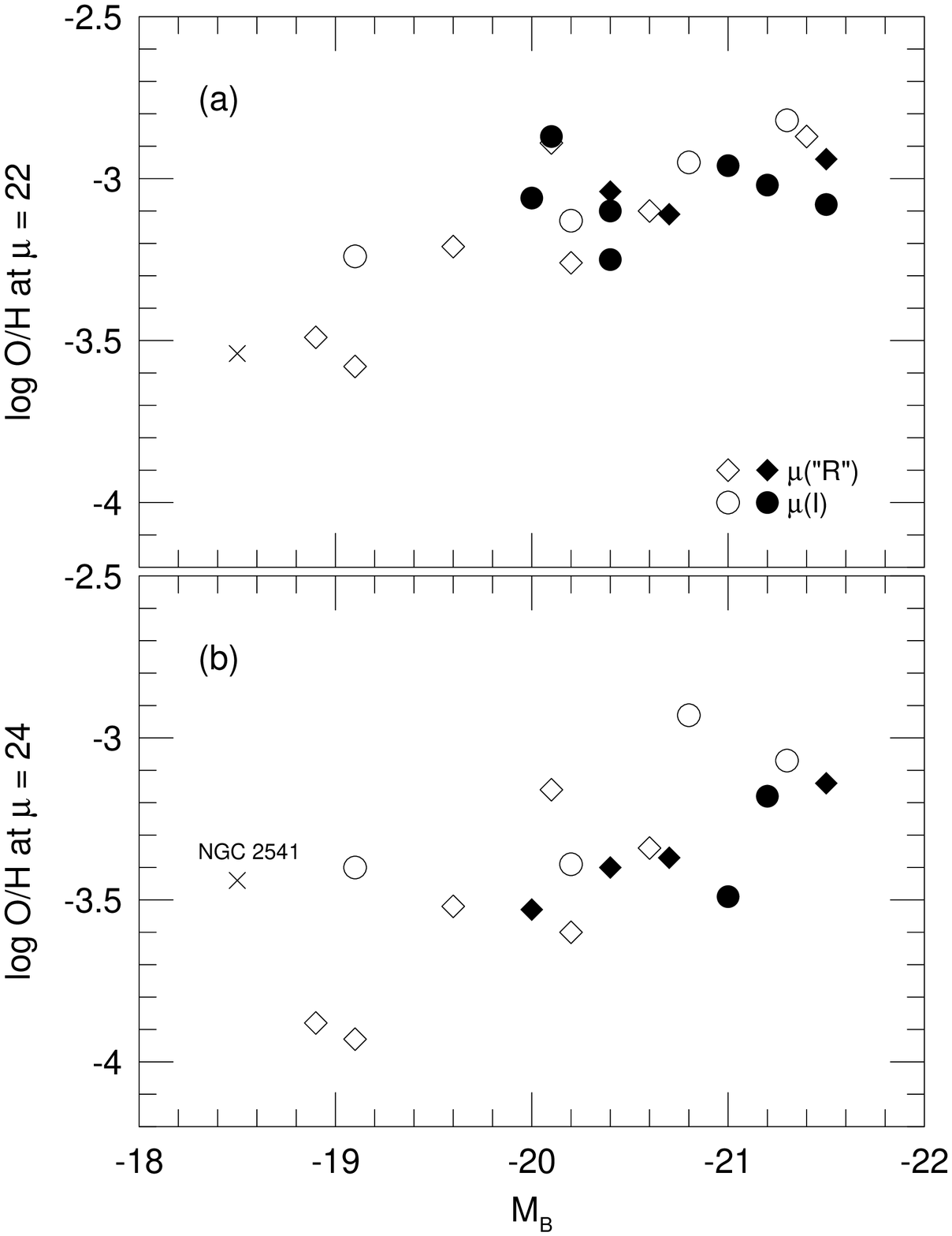}} \hfill
\parbox[b]{\hsize}{
\caption[]{O/H at fixed value of galaxy surface brightness vs. $M_B$. Top: 
abundances at 22 mags arcsec$^{-2}$. Bottom: abundances 
at 24 mag arcsec$^{-2}$. More luminous spirals have higher abundances at 
a fixed surface brightness. Open symbols are from Garnett et al. (1997a); 
filled symbols represent additional data obtained from the literature. }}
\end{figure}

The uniformity of abundance gradients as a function of scale length 
suggests a close correlation between metallicity and disk surface 
brightness. Indeed, McCall (1982) and Edmunds \& Pagel (1984) noted 
a remarkably tight correlation between O/H and disk surface brightness 
for late-type spirals. This has provided part of the motivation for 
models of self-regulated star formation, in which the radiation and 
mechanical energy produced by stars feeds back into the surrounding 
ISM and acts to inhibit further star formation. Models of this kind 
have been explored by Phillips \& Edmunds (1991) and Ryder (1995), 
and appear to do a good job of reproducing the trends of both star 
formation rate and O/H with surface brightness. One caveat is that
the interaction of the stellar energy output with the ISM is still 
poorly understood. Viscous disk models also provide a possible mechanism 
to tie the abundances to the underlying surface density distribution.

Edmunds \& Pagel (1984) also noted that early-type spirals do not follow 
the same O/H-surface brightness correlation as the late types.
Garnett et al. (1997a) put this on a more quantitative 
basis. Figure 13 displays the characteristic metallicity at two fixed 
values of disk surface brightness for a sample of spirals having either 
I- or R-band surface photometry; these bandpasses presumably sample the 
light from the old disk population better than B. The figure shows that 
metallicity-luminosity correlation appears to hold at all values of
surface brightness across spiral disks. This result argues for two
modes of enrichment in disk galaxies: a local mode, in which the
metallicity is connected to the local mass density, and a global 
mode, in which an entire galaxy is enriched in a manner dependent
on its total mass. One can imagine a global enrichment event which
raises the metallicity of a galaxy to some level which depends on
total mass, followed by sequential local enrichment which follows
the mass density distribution. One caveat is that $M/L$, and thus
the mass surface density at a given surface brightness, may vary 
systematically along the luminosity sequence in Figure 13. A more
comprehensive study of mass surface density and gas fraction along
this sequence should prove enlightening.


\subsection{Barred Spirals}

\begin{figure}[t!]\label{fig_gvsbar}
\resizebox{\hsize}{!}{\includegraphics{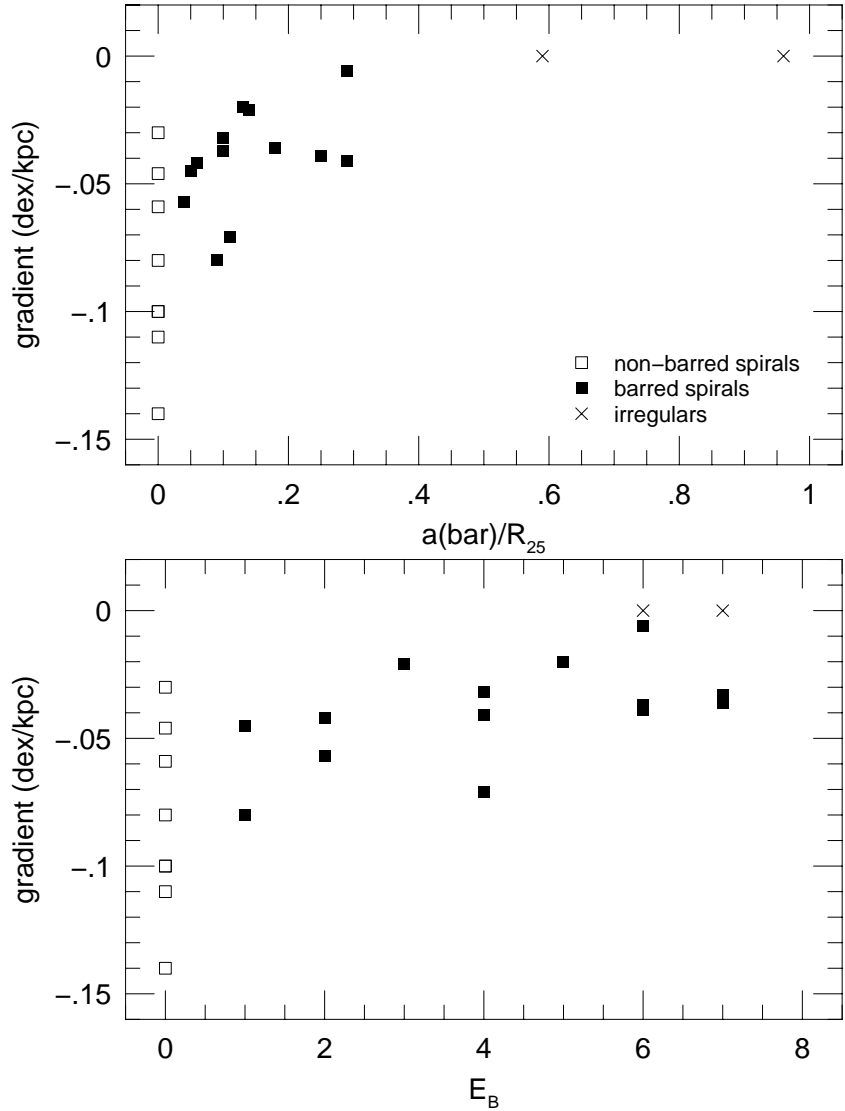}} \hfill
\parbox[b]{\hsize}{
\caption[]{{\it Top:} O/H gradient per kpc vs. bar length $a$ relative to the
photometric radius $R_{25}$. Filled squares are barred spirals, unfilled
squares non-barred spirals, and crosses are irregular galaxies. Bottom: 
O/H gradient vs. bar strength $E_B$ = 10(1$-b/a$), where $b/a$ is the
ratio of bar minor and major axes. Plot adapted from Martin \& Roy 1994. }}
\end{figure}

Bars are interesting because the gravitational potential of a bar is 
expected to induce a large-scale radial gas flow, possibly through
radiative shocking and subsequent loss of angular momentum as the
gas passes through the bar (Barnes 1991). The radial flow could
significantly alter the metallicity distribution by mixing in gas
from outer radii, thus weakening composition gradients, and is often
argued to be the means to fuel the nuclei of active galaxies. The 
evidence so far accumulated indicates that barred spirals generally 
do have shallower composition gradients than weakly-/non-barred spirals 
(ZKH; Martin \& Roy 1994). Martin \& 
Roy (1994) have argued that the slope of the composition gradient
correlates with both bar length and bar strength, defined as the 
ratio of bar length to width. This is illustrated in Figure 14, 
which shows the slope of the O/H gradient per kpc versus bar length
$a$ relative to the photometric radius ({\it top panel}) and versus
bar strength $E_B$ = 10(1$-b/a$), where b is the bar width. Non-barred,
barred, and irregular galaxies from the Martin \& Roy sample are 
distinguished by different symbols, and the gradients have been 
adjusted for new distances to the galaxies. The plots show that (1)
the non-barred spirals show a wide range of values for gradient slopes,
although on average they are steeper than those for barred spirals;
and (2) the O/H gradients for the barred spirals tend to get shallower
with increasing bar length and bar strength. On the other hand, the
trend depends how the gradients are scaled; if one plots the O/H
gradient per unit scale length instead, the correlations disappear.

Curiously, some barred spirals (e.g., NGC 1365; Roy \& Walsh 1997) show 
an O/H gradient within the bar, with flattening only outside the bar. 
One way to explain this is if strong star formation occurs in the
bar, building up abundances faster than the radial flow can homogenize
them (Friedli, Benz, \& Kennicutt 1994). 

\subsection{Spiral Bulges}

Abundances in bulges generally are obtained the same way as for
ellipticals, by spectroscopy and modeling of absorption line indices
from the integrated stellar population, and so suffer from the same
uncertainties. Most of the line indices in the Lick system vary 
in the same way with age and metallicity, and so poorly distinguish 
between the two in model grids. A few, such as H$\beta$, Mg $b$, 
Fe5270, and Fe5335, do provide some ability to separate age and
metallicity, and have been used to obtain estimates of ages, [Fe/H],
and [Mg/Fe] in spheroidal systems (Worthey 1998).  

Most work on abundances in bulges have come from spectroscopy of the 
central regions. Jablonka, Martin, \& Arimoto (1996) found that Mg$_2$ 
correlated with both bulge luminosity and stellar velocity dispersion 
in spirals with T = 0-5, but that the Fe5270 feature did not. Comparing 
with a grid of synthetic spectra with non-solar [$\alpha$/Fe] they 
inferred that [Mg/Fe] increased with bulge luminosity as well. Similar 
results are obtained for ellipticals (Worthey 1998). Maps of spectral 
line indices obtained with integral field units (Peletier et al. 1999, 
de Zeeuw et al. 2002) show hints that Mg/Fe varies with radius, although 
only a small number of galaxies have been analyzed so far. 

High-resolution spectroscopy of red giants in our own Galactic bulge
recommends caution in interpreting line indices in integrated spectra
of ellipticals and bulges. Giants in the Baade's Window region show
high $\alpha$/Fe ratios and a mean [Fe/H] $\approx$ $-$0.25, similar
to [Fe/H] for solar neighborhood stars (McWilliam \& Rich 1994).
The mean metallicity is lower
by about 0.3 dex compared to low-resolution spectroscopic and photometric
determinations.
This has several implications. Enhanced Ti/Fe ratios
make the spectral types of the bulge giants later for the same IR colors 
compared to stars with solar abundance ratios. Enhanced $\alpha$/Fe
alter both stellar line indices and the location of isochrones in 
population synthesis models, which have been computed using solar
abundance ratios so far. Thus, metallicities derived for spheroids
may be overestimated by a factor of two or so. Most population synthesis
Trager et al. (2000a,b) attempt to correct the models for the effects
of non-solar element abundances ratios. Such corrections are non-trivial,
as both isochrones and line strengths are affected. No such corrections
have been applied to bulges yet.

The formation of bulges is still mysterious. The candidate mechanisms
are: monolithic collapse with rapid star formation (Eggen, Lynden-Bell,
\& Sandage 1962); mergers of roughly equal mass objects in hierarchical
clustering models for galaxy formation (Baugh, Cole, \& Frenk 1996: 
Kauffmann 1996); and secular growth from disk material, for example by 
mass transfer via bars (Combes et al. 1990; Hasan, Pfenniger, \& Norman 
1993). High $\alpha$/Fe (that is, higher than the solar ratio) would 
tend to favor the models with rapid bulge formation from relatively 
metal-poor gas, because most of the Fe is expected to come from Type
Ia supernovae. Solar or less $\alpha$/Fe would tend to favor secular 
evolution from already enriched material, or star formation extended 
over times cales greater than 1 Gyr. The correlation of Mg/Fe with
bulge luminosity and velocity dispersion suggests that a mix of formation
mechanisms are at work; moreover, Andredakis et al. (1995) have found 
that bulge structure parameters correlate with Hubble type and bulge
luminosity, such that large, luminous bulges tend toward $R^{1/4}$ 
profiles similar to ellipticals, while small bulges tend to have 
exponential profiles (although de Jong 1996 argues that bulges have
exponential profiles in general). The trends in Mg/Fe and bulge shape
together suggest that large bulges formed rapidly at early times,
while small bulges may have formed more slowly via secular processes.

\subsection{Cluster Spirals and Environment}

Environment and interactions appear to play a significant role in the 
evolution of galaxies, particularly in dense environments. Interactions 
with satellites may be responsible for the significant fraction of 
lopsided spiral galaxies (Rudnick \& Rix 1998). Disk asymmetry may 
affect the inferred spatial distribution of metals in the interstellar 
gas. For example, Kennicutt \& Garnett (1996) noted an asymmetry in O/H 
between the NW and SE sides of M101, which may be related to the asymmetry 
in the structure of the disk. Zaritsky (1995) found a possible correlation 
between disk $B-V$ and the slope of the O/H gradient such that bluer 
galaxies tended to have steeper gradients. He suggested that accretion 
of metal-poor, gas-rich dwarf galaxies in the outer disk could steepen 
abundance gradients and make the colors bluer through increased star 
formation.  On the other hand, the trend may simply reflect the fact 
that spirals with steep metallicity gradients tend to be lower-luminosity 
late Hubble types with bluer colors on average. 

Rich clusters offer a variety of galaxy-galaxy and galaxy-ICM interactions.
The cluster environment certainly affects the morphology of galaxies 
(Dressler 1980). It is also known that spirals near the center of rich
clusters, for example, the Virgo cluster,  show evidence for stripping 
of H~I, especially from the outer disks (Warmels 1988; Cayatte et al.
1994). Such stripping is inferred to result from interaction of the 
galaxy ISM with the hot X-ray intracluster gas. The degree of H~I 
stripping correlates with projected distance from the cluster core,
although in Virgo the molecular content appears to be not affected
(Kenney \& Young 1989). 

If field galaxies evolve through continuing infall of gas (Gunn \&
Gott 1972), then the truncation of H~I disks in cluster spirals 
should have an effect on the chemical evolution. Specifically, 
infall of metal-poor gas reduces the metallicity of the gas at a
given gas fraction. Truncation of such infall should then cause 
the chemical evolution of cluster spirals to behave more like the
simple closed box model, and thus should have higher metallicities
than comparable field spirals. This idea has led to several studies
of oxygen abundances in Virgo spirals. The largest study so far is
that of Skillman et al. (1996), who obtained data for nine Virgo
spirals covering the full range of H~I deficiencies. The results
indicate that cluster spirals with the largest H~I deficiencies 
have higher O/H abundances than field spirals with comparable 
$M_B$, rotation speeds, and Hubble types, while spirals with only
modest or little H~I stripping have abundances comparable to those
of similar field spirals (Skillman et al. 1996, Henry et al. 1994,
1996). The samples studied so far have been small, and one must
worry about possible systematic errors in abundance caused by the
lack of measured electron temperatures. This is an area that could
benefit from further study with larger galaxy samples and more 
secure abundance measurements.

\section{Element Abundance Ratios in Spiral and Irregular Galaxies}

The abundance ratios of heavy elements are sensitive to
the initial mass function (IMF), the star formation history, and variations
in stellar nucleosynthesis with, e.g., metallicity. In particular, comparison
of abundances of elements produced in stars with relatively long lifetimes
(such as C, N, Fe, and the s-process elements) with those produced in 
short-lived stars (such as O) probe the star formation history. Below, I
review the accumulated data on C, N, S, and Ar abundances (relative to O) 
in spiral and irregular galaxies, covering two orders of magnitude in 
metallicity (as measured by O/H). The data are taken from a variety of 
sources on abundances for H~II regions in the literature. 

\subsection{Helium}

Helium, the second most abundant element, has significance for cosmology
and stellar structure. Most $^4$He was produced in the Big Bang, and the
primordial mass fraction $Y_p$ is a constraint on the photon/baryon
ratio and thus on the cosmological model. The He mass fraction also
affects stellar structure, but He is difficult to measure in stars and
so must be inferred from other measurements. On the other hand, He~I 
recombination lines are relatively easy to measure in H~II regions,
and so a large amount of data is available on He/H in ionized nebulae.

A great deal of effort has been spent in determining $Y_p$, and is covered 
in Gary Steigman's contribution, so I will be brief on this aspect. Peimbert 
\& Torres-Peimbert (1974) initiated the current modern study of $Y_p$ by 
making the simple assumption that the He mass fraction varies linearly with 
metallicity (or O/H), and thus used measurements of abundances in H~II 
regions with a range of O/H to extrapolate to the pre-galactic He abundance 
at O/H = 0. Today there is very high signal/noise data on He abundances
in approximately 40 metal-poor dwarf irregulars with O/H ranging from 2\%
to 10\% solar, so the extrapolation to O/H = 0 can be estimated to high
statistical precision. 

The good news is that the best current estimates of $Y_p$ agree to within
5\%. This is amazing agreement for measurements derived from spectroscopy
of distant galaxies, so we should all feel proud. Nevertheless, the 
differences in $Y_p$ estimates are a source of consternation for theory
of cosmological nucleosynthesis, as the two largest studies obtain 
values for $Y_p$ which disagree at the 4-5$\sigma$ significance level: 
Olive, Skillman \& Steigman (1997) derived $Y_p$ = 0.234$\pm$0.003, while 
Izotov \& Thuan (1998) derived $Y_p$ = 0.244$\pm$0.002 (statistical 
uncertainties only for both studies), from similar-sized H~II region 
samples. Depending on which estimate is considered most reliable, 
$Y_p$ either agrees with the best current estimate of D/H under 
standard Big Bang nucleosynthesis (for $Y_p$ = 0.244, or it does not. 

\begin{figure}[t!]\label{fig_fkghelium}
\resizebox{\hsize}{!}{\includegraphics{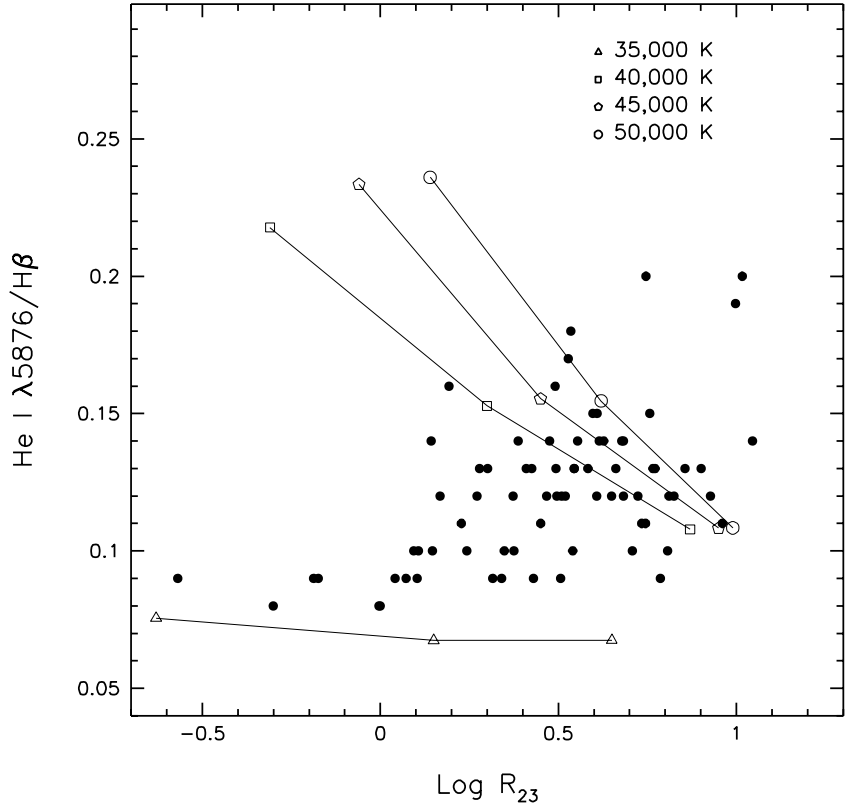}} \hfill
\parbox[b]{\hsize}{
\caption[]{I(He~I $\lambda$5876)/I(H$\beta$) vs. $R_{23}$ from 
spectroscopy of H~II regions in spiral galaxies (Bresolin et al. 
1999), showing how the He~I line strengths decrease for low $R_{23}$
(high O/H). Overplotted are trends obtained from photoionization
models with various values of $T_{eff}$ for the ionizing stars,
assuming that He varies with metallicity as $\Delta$Y/$\Delta$ = 2.5.
}}
\end{figure}

At present the battleground for $Y_p$ is focused on sources of systematic
error, and these are likely to yield the greatest improvements in He 
measurements, rather than measuring more data points. The areas that 
need work are:
\begin{itemize}
\item Corrections for He~I absorption by the underlying OB association.
These are fairly uncertain and affect the He~I line ratios as well as
the total He abundance. B main sequence and supergiants have the largest
He~I line strengths, and so need special attention; the B supergiant
contribution is likely to be affected by stochasticity and uncertain
stellar evolution.
\item Density effects. He~I lines in the optical spectrum, particularly 
the triplets, are subject to collisional excitation because the 2 $^3$S
level is metastable. The contribution of collisional excitation depends
on both electron density and electron temperature. It is debated whether
electron densities typically derived from [S II] line ratios are 
appropriate for He~I. Detailed studies of density structure in a few  
good H~II regions would provide useful information on this.
\item Radiative transfer. Again, because the 2 $^3$S level is metastable,
transitions decaying into this level can build up large optical depths
in H~II regions. This leads to redistribution of line ratios among the
triplets. This problem is coupled to the collisional excitation problem.
\item Corrections for neutral He. He$^0$ can not be observed directly
in H~II regions; the He$^0$ fraction must be inferred from ionization
models. Since the ionization energy of He is 24.6 eV He$^0$ can exist
in the H$^+$ zone. However, for ionizing radiation field with an
effective temperature $T_{eff}$ $>$ 40,000 K the He$^+$ Str\"omgren
radius is nearly coincident with the H$^+$ radius. This problem is
largely solved by observing only high-ionization H~II regions, with
O$^+$/O $<$ 0.15. Nevertheless, even in this case the He$^0$ corrections
can be 1-2\%, either positive or negative. 
\item Collisional excitation of H~I. In high temperature metal-poor H~II 
regions, collisional excitation of H$\alpha$ could be significant (of 
order 5\% or so). H~I collisional excitation affects the He/H ratio and 
interstellar reddening estimates. The effect is very sensitive to the 
fraction of H$^0$ present in the highly-ionized zone, which is very 
uncertain. 
\end{itemize}
Each of these error sources contribute perhaps 1-2\% to the uncertainty
in derived He abundances, but it is how they sum that determines the
systematic error, which is not fully understood.

Helium abundances in spiral galaxies are less well-determined, because
of more uncertain electron temperatures. The He abundance does have
an effect on ionization structure, so it is of interest to know how
He/H varies with metallicity in the inner disks of spirals. Does He/H
continue to rise linearly with metallicity as in the metal-poor galaxies,
or does it level off? This may be difficult to determine from H~II
region spectroscopy, as there appears to be a drop in the He~I line
strengths in the inner disks of spirals (Bresolin, Kennicutt, \& Garnett
1999), contrary to what one would expect from ionization models with
rising or even constant He/H (Figure 15). The low He~I 5876 line strengths
in the metal-rich regions observed so far are lower than expected even 
for primordial He/H. The trend of decreasing He~I line strength is
best explained if ionizing clusters in H~II regions with metallicity 
above solar have radiation fields with characteristic temperatures
of about 35,000 K. This is the strongest evidence available for a
possible variation in the upper limit of the massive star IMF.

\subsection{Carbon}

Carbon abundances in H~II regions have been difficult to determine
with precision because the important ionization states (C$^+$, C$^{+2}$)
have no strong forbidden lines in the optical spectrum. Only the UV
spectrum shows collisionally excited lines from both C~II and C~III.
A number of studies of carbon abundances in extragalactic H~II regions
were made with $IUE$ (e.g., Dufour, Shields, \& Talbot 1982; Peimbert,
Pe\~na, \& Torres-Peimbert 1986; Dufour, Garnett \& Shields 1988), but 
for the most part the $IUE$ observations suffered from low signal/noise
and uncertainties due to aperture mismatches between UV and optical
spectra.

\begin{figure}[t!]\label{fig_CO}
\resizebox{\hsize}{!}{\includegraphics{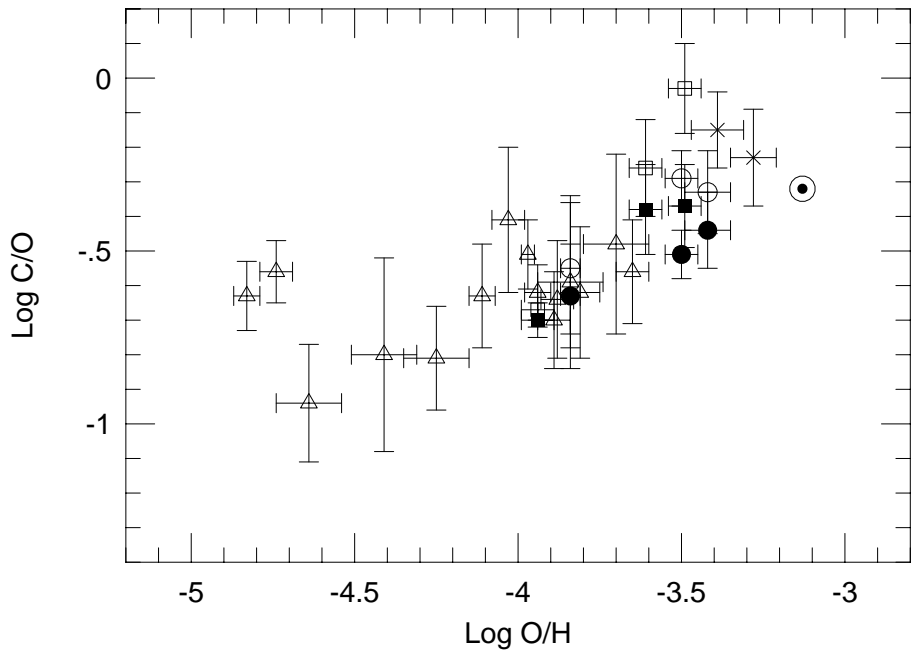}} \hfill
\parbox[b]{\hsize}{
\caption[]{C/O abundance ratios (by number) from spectroscopy of H~II regions 
in spiral and irregular galaxies (Garnett et al. 1995a, 1997b, 1999). {\it Open 
symbols}: irregular galaxies; {\it filled symbols}: spiral galaxy H~II regions. 
}}
\end{figure}

\begin{figure}[t!]\label{fig_comodcomp}
\resizebox{\hsize}{!}{\includegraphics{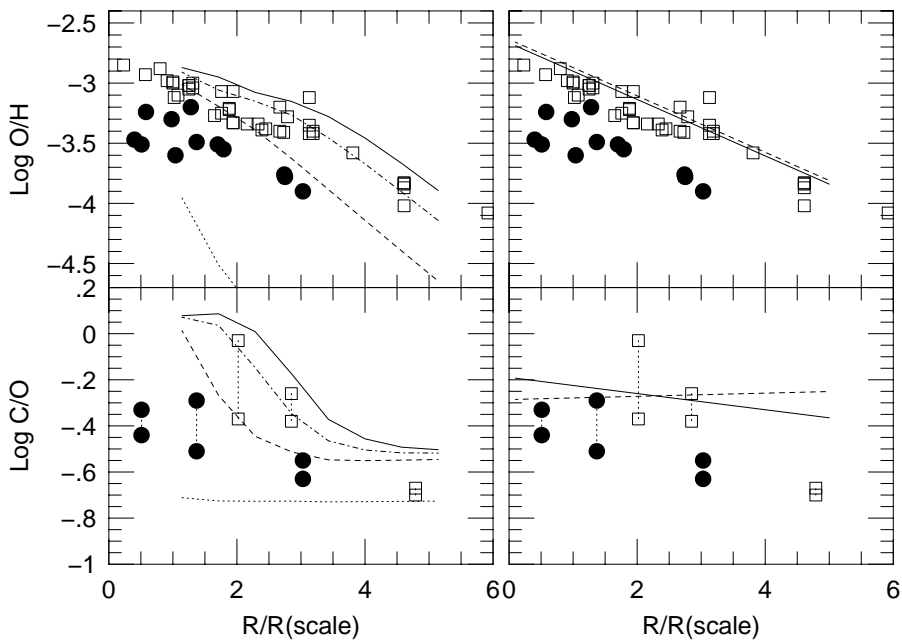}} \hfill
\parbox[b]{\hsize}{
\caption[]{C/O and O/H gradients in M101 (open squares) and NGC 2403 (filled
circles) plotted vs. radius normalized to the disk scale length (Garnett et 
al. 1999). {\it Left panels:} Galactic chemical evolution models from Carigi
1996, using massive star yields from Maeder 1992. The curves show the 
O/H and C/O gradients at different times, from 0.5 Gyr ({\it dotted curves})
to 13 Gyr ({\it solid curves}).
{\it Right panels:} Galactic chemical evolution models from G\"otz \&
K\"oppen 1992 ({\it dashed line}) and Moll\'a et al. 1997 ({\it solid
lines}) using massive star yields with no stellar winds. All models use
the same intermediate-mass star yields from Renzini \& Voli 1981.
}}
\end{figure}

The higher UV sensitivity of {\it HST} offered greatly improved measurements
of UV emission lines from [C~II] and C~III], plus the opportunity to scale the
C lines directly to [O~II] and O~III] lines in the UV, tremendously reducing
the uncertainties due to reddening corrections and errors in $T_e$ (Garnett
et al. 1995a, 1999; Kobulnicky \& Skillman 1998). The most recent data for C/O 
as a function of O/H in dwarf irregular and spiral galaxies from {\it HST} 
measurements are displayed in Figure 16. Some C (and O) is expected to be 
depleted onto interstellar dust grains. Sofia et al. (1997) showed that the 
gas-phase C abundance varies little with physical conditions in the local 
neutral ISM, suggesting a constant fraction of C in dust everywhere. They 
infer that C is depleted by about 0.2 dex. O should be depleted by no more 
than $\approx$ 0.1 dex everywhere (Mathis 1996). Thus, it is likely that our 
C/O values should all be increased by 0.1-0.2 dex, but we do not expect
any systematic variation in the fractional depletions with metallicity. 

Figure 16 shows a trend of steeply increasing C/O for log O/H $>$ $-$4. 
This is in agreement with observations of C/O in disk stars in the
Galaxy (Gustafsson et al. 1999). 
The C/O ratios in the most metal-poor galaxies are consistent with the 
predictions for massive star nucleosynthesis by Weaver \& Woosley (1993; 
hereafter WW93) for their best estimate of the
$^{12}C(\alpha,\gamma)^{16}O$
nuclear reaction rate factor. On the other hand, the amount of contamination 
by C from intermediate mass stars is poorly known in these galaxies.

The notable trend in Figure 16 is the apparent `secondary' behavior
of C with respect to O, despite the fact that C (i.e., $^{12}$C) is
primary. Tinsley (1979) demonstrated that such variations can be 
understood as the result of finite stellar lifetimes and delays in
the ejection of elements from low- and intermediate-mass stars. If
C is produced mainly in intermediate-mass stars, then the enrichment
of C in the ISM is delayed with respect to O, which is produced in 
high-mass stars. 

At the same time, C is also produced in high-mass stars, with a 
production yield that is fairly uncertain. In stars without mass
loss, the relative yield of C with respect to O in massive stars is
smaller than the solar ratio (WW93), which would 
demand that most C come from intermediate-mass stars. Maeder (1992),
however, showed that stellar mass loss can affect the yields of 
C and O from massive stars. The effect of such mass loss is to 
remove He and C from the massive stars before they can be further 
processed into O. If the mass-loss rates depend on radiative opacity,
and thus on metallicity, then the yields of C and O will depend on
metallicity, with the C yield increasing with $Z$ at the expense of O. 

Figure 17 shows the data for the spiral galaxies M101 and NGC 2403 with 
the predictions of two sets of chemical evolution models overlaid. The 
left panels show a sequence of Galactic chemical evolution models using 
massive star nucleosynthesis models including stellar winds from Maeder 
(1992); the right panels shows two other Galactic chemical evolution 
models derived with massive star yields computed assuming no stellar 
mass loss. All of the models use the same intermediate-mass star yields.
Although all of the models reproduce the O/H gradients reasonably well,
only the models with Maeder yields seem able to reproduce the steep C/O 
gradients observed - with the caveat that these models were not tailored 
for the two galaxies in question. Comparison of solar neighborhood models 
with the observations of stars also tend to favor the nucleosynthesis 
models that take into account metallicity-dependent mass loss for massive 
stars (e.g., Prantzos et al. 1994; Carigi 2000). 

The big uncertainties in all of this revolve around the theoretical
yields. Problem number one is the 
$^{12}C(\alpha,\gamma)^{16}O$ reaction rate,
which is still highly uncertain (Hale 1998). Problem two is uncertainty
in convective mixing. $^{16}$O is produced by $\alpha$ captures onto 
$^{12}$C during helium burning. Mixing of fresh He into the convective
zone can turn C into O rapidly (Arnett 1996, pp. 223-229). Finally, mass 
loss rates for stars in various evolutionary states and metallicities are 
also still uncertain. For intermediate-mass stars, differences in mixing
and treatment of thermal pulses affect the C yields. The most recent
models for nucleosynthesis in intermediate-mass stars still show
large discrepancies in yields (Portinari et al. 1998; van den Hoek
\& Groenewegen 1997; Marigo et al. 1996, 1998). Until these problems
are solved or we have empirically-derived C yields for stars of various
masses, it will be difficult to reliably interpret the abundance trends.

\subsection{Nitrogen }

Nitrogen abundances in extragalactic H~II regions are almost entirely
derived from optical [N~II] lines alone, because the other important
species, N III, has strong emission lines only in the UV and FIR. 
Photoionization models generally predict that N$^+$/O$^+$ = N/O under
most conditions. Nevertheless, IR measurements of [N~III]/[O~III] in
Galactic H~II regions consistently find a steeper N/O gradient than
that obtained from optical measurements of [N~II]/[O~II] (Lester et al.
1987; Mart\'\i n-Hern\'andez et al. 2002). This suggests that ionization
corrections may be important (Garnett 1990). Direct comparison between
[N~II] and [N~III] measurements in H~II regions with varying properties
is needed to understand the nitrogen ionization balance, so that the
variation in N with metallicity can be studied accurately.
Comparison with measurements in stars is also helpful, and it should
be noted that measurements of abundances in B stars (e.g. Rolleston et
al. 2000, Korn et al. 2002) yield O and N abundances in good
agreement with the values for H~II regions in the Milky Way and the
LMC and the Local Group spiral M33 (McCarthy et al. 1995; Monteverde 
et al. 1997). 

\begin{figure}[t!]\label{fig_NO}
\resizebox{\hsize}{!}{\includegraphics{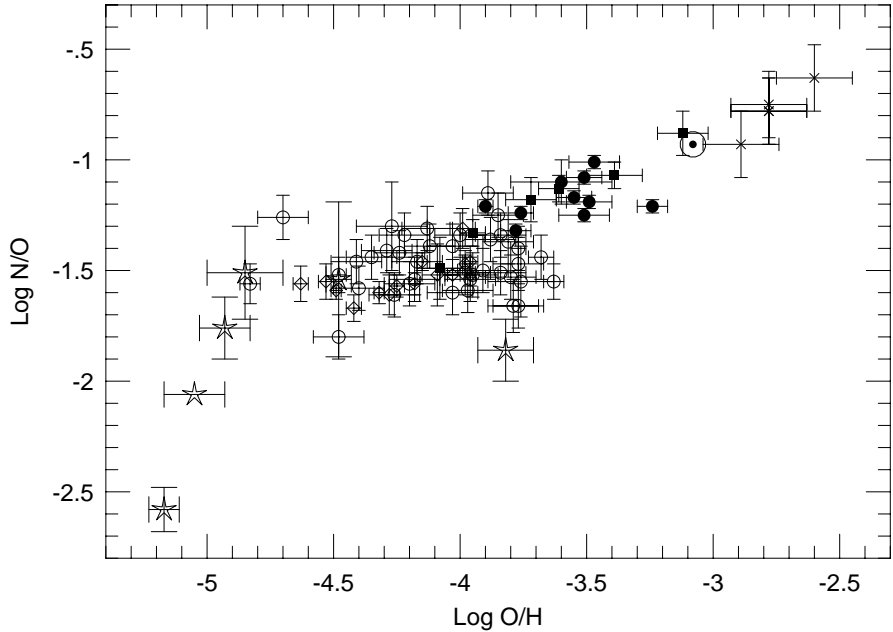}} \hfill
\parbox[b]{\hsize}{
\caption{N/O abundance ratios in spiral and irregular galaxies. {\it Open 
circles}: Garnett 1990; {\it open diamonds}: Thuan et al. 1995; 
{\it filled circles}: Garnett et al. 1999; {\it filled squares}: Garnett and 
Kennicutt 1994, Torres-Peimbert et al. 1989; {\it plus signs}: D\'\i az et 
al. 1991; {\it stars}: high-redshift absorption line systems from Lu, 
Sargent, and Barlow 1998. Note the very low N/O in some high-redshift systems.}}
\end{figure}

With this uncertainty in mind, Figure 18 shows how N/O varies with O/H 
from optical spectroscopy of H~II regions in spiral and irregular galaxies. 
It has been known for some time that N seems to have two components: one 
component which follows O in a fixed ratio (log N/O $\approx$ $-$1.5) for 
log O/H $<$ $-$3.7, as inferred from the constant N/O vs. O/H in metal-poor 
dwarf irregular galaxies (open circles), and a second component that increases 
faster than O at higher O/H as seen in spiral galaxies (filled circles and
squares, crosses). The second component is produced via the classical CNO 
cycle during hydrogen burning in stars and requires the presence of C and 
O in the star from birth (``secondary N''), while the first component is 
postulated to come from the CN cycle on freshly-synthesized C (from He-burning) 
which has been convectively ``dredged-up'' into a hot H-burning zone at the 
base of the convective envelope, and does not require an initial seed of C 
or O (hence, ``primary'' N). The latter process is most commonly thought to 
occur in the asymptotic giant branch (AGB) stage of intermediate mass stars 
(Iben \& Truran 1978), but has been found to occur in models of massive stars 
with increased convective overshooting or rotationally-induced mixing (e.g. 
Langer et al. 1997). It has been unclear which primary N source accounts 
most for the constant N/O in the dwarf galaxies. The massive star primary 
source does not appear to produce enough N to yield N/O $\approx$ 0.03. 
For the lower mass stars, the various AGB model calculations give rather 
discrepant results for the production of N (see Forestini \& Charbonnel
1997; van den Hoek \& Groenwegen 1997; Marigo 2001). The N production
during the third dredge-up is very sensitive to the assumptions that
determine the boundary of the convective zone and the overshoot. This
is a highly hydrodynamic problem including explosive thermal pulse events
and is difficult to model at present (Lattanzio 1998). It is likely that
this will continue to be an important topic of study for the near future.

Some insight may be found by examining more distant objects. Recent studies 
of high-redshift Lyman-alpha absorption systems (plotted as stars in Figure 
18) have found objects with N/S, N/Si, and N/O ratios much lower than in 
the dwarf galaxies (Pettini, Lipman \& Hunstead 1995; Lu, Sargent, and 
Barlow 1998). A wide range in inferred N/O is seen in the DLAs, but the
lowest values are as much as a factor 10 smaller than the average for
irregular galaxies. 
Although S and Si column densities are derived from S II and Si II, 
which can coexist with both ionized and neutral gas, ionization effects 
appear to insufficient to account for low N/S and N/Si (Vladilo et al. 2001).
The results are consistent with the idea that the DLAs represent lines of 
sight through very young galaxies, with an age spread of a few hundred Myr, 
the timescale for enrichment of N from AGB stars. The higher N/O ratios
seen in irregular galaxies would then be largely the product of AGB stars.

\begin{figure}[t!]\label{fig_NeO}
\resizebox{\hsize}{!}{\includegraphics{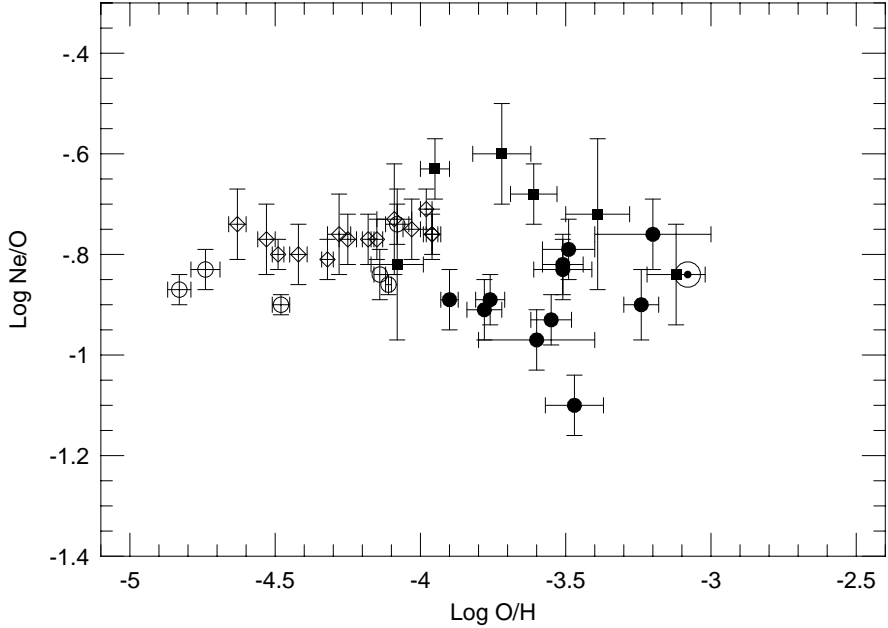}} \hfill
\parbox[b]{\hsize}{
\caption{Ne/O abundance ratios in spiral and irregular galaxies. Symbols 
are the same as in Figure 18. }}
\end{figure}

Some scatter is seen in N/O for the more metal-rich dwarf galaxies 
(Kobulnicky \& Skillman 1998). This may be the result of localized
enrichment by Wolf-Rayet stars. The most metal-poor dwarf galaxies
seem to show very little scatter in N/O (Thuan et al. 1995). It is 
possible that this may simply reflect small number statistics (dwarf 
galaxies with log O/H $<$ $-$4.5 and bright H~II regions are rare). 
It is also possible to understand these galaxies if they are 
relatively old systems that experienced an episode of star formation
in the past which enriched them to their present composition, and
are experiencing a new starburst event after a long quiescent period.

\subsection{Neon, Sulfur and Argon}

Neon, sulfur, and argon are products of the late stages of massive
star evolution. $^{20}$Ne results from carbon burning, while S and
Ar are products of O burning. As they are all considered part of the
$\alpha$-element group, their abundances are expected to track 
O/H closely.

Neon abundances in extragalactic H~II regions are derived primarily 
from optical measurements of [Ne~III], although spacecraft measurements 
of the IR [Ne~II] and [Ne~III] fine-structure lines are becoming 
available. A representative sample of Ne/O values for H~II regions 
with measured $T_e$ in spiral and irregular galaxies is shown in 
Figure 19. The scatter increases for the H~II regions with higher
O/H because of the more uncertain electron temperatures. It is 
apparent that the Ne abundance tracks O quite closely, in agreement
with results from planetary nebulae (Henry 1989). 

\begin{figure}[t!]\label{fig_SArO}
\resizebox{\hsize}{!}{\includegraphics{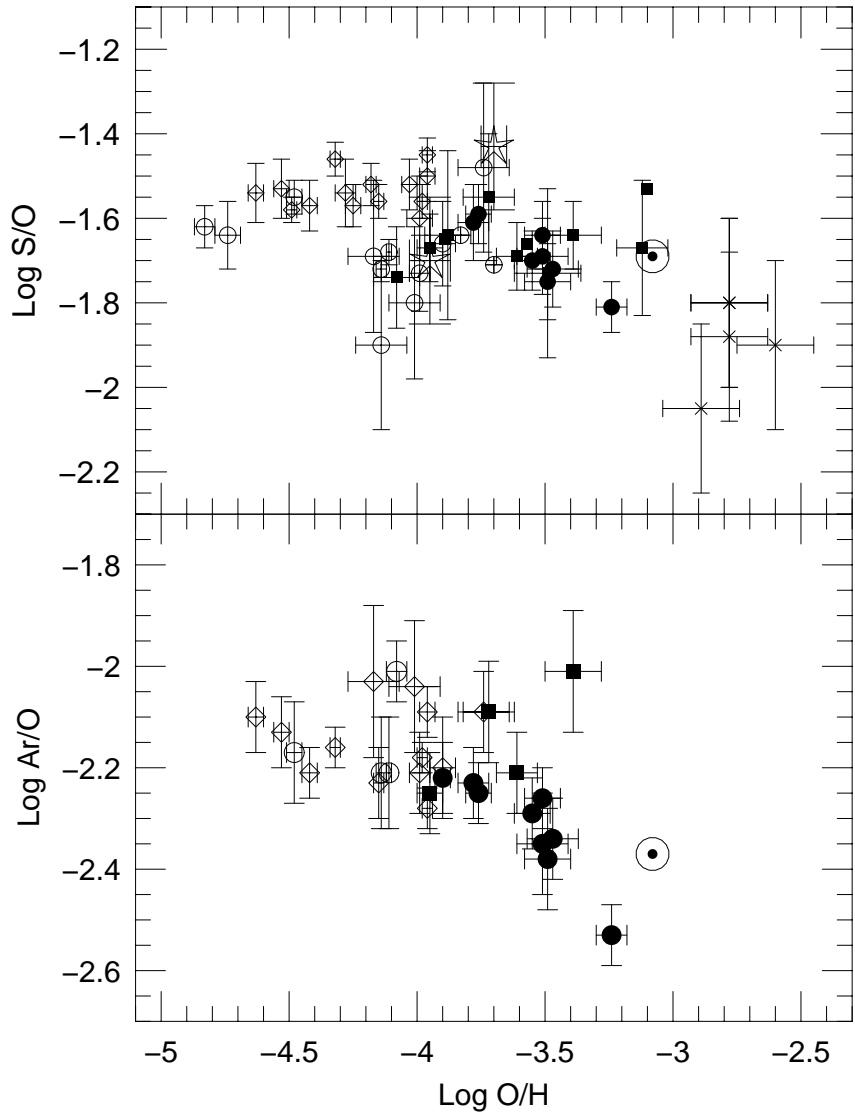}} \hfill
\parbox[b]{\hsize}{
\caption{{\it Top:} S/O abundance ratios in spiral and irregular 
galaxies. {\it Bottom:} Ar/O abundance ratios in spiral and 
irregular galaxies. Symbols are the same as in Figure 18. 
}}
\end{figure}

Figure 20 shows data for sulfur and argon.
For log O/H $<$ $-$3.5, S/O and Ar/O are essentially constant with O/H, 
and fall within the range predicted by WW93. 
For log O/H $>$ $-$3.5, however, there is evidence for declining S/O and 
Ar/O as O/H increases. The cause of the decline is not clear. It is
possible that the ionization corrections for unseen S$^{+3}$ have been
underestimated in the more metal-rich H~II regions. More observational 
study, especially IR spectroscopy, of metal-rich H~II regions is needed, 
to rule out ionization or excitation effects. 

Because S and Ar are produced close to the stellar core, the yields of 
S and Ar may be sensitive to conditions immediately prior to and during 
the supernova explosion, such as explosive processing or fall-back onto 
the compact remnant (WW93). If real, however, the 
declining S/O and Ar/O cannot be accounted for by simple variations in 
the stellar mass function and hydrostatic nucleosynthesis (Garnett 1989); 
some variation in massive star and/or supernova nucleosynthesis at high 
metallicities (perhaps due to strong stellar mass loss) may be needed. 

\subsection{Other Elements}

Few other elements have been measured systematically in extragalactic
H~II regions, over a wide range of O/H. 

Silicon can be measured in H~II regions through the UV Si~III]
doublet at 1883, 1892 \AA. Figure 21 shows Si/O measurements from
Garnett et al. (1995b) for a small sample of metal-poor galaxies,
along with data for several samples of B stars and the Orion nebula.
Si/O appears to be roughly constant but smaller than the average
for the Sun and the solar neighborhood B stars. Silicon is certainly
depleted onto grains in the ISM, and the results in Figure 21 are
consistent with a Si depletion of about $-$0.2 to $-$0.4 dex in
the H~II region sample. This is probably appropriate for not very
dense, ionized gas (Sofia, Cardelli, \& Savage 1994). 

\begin{figure}[t!]\label{fig_SiO}
\resizebox{\hsize}{!}{\includegraphics{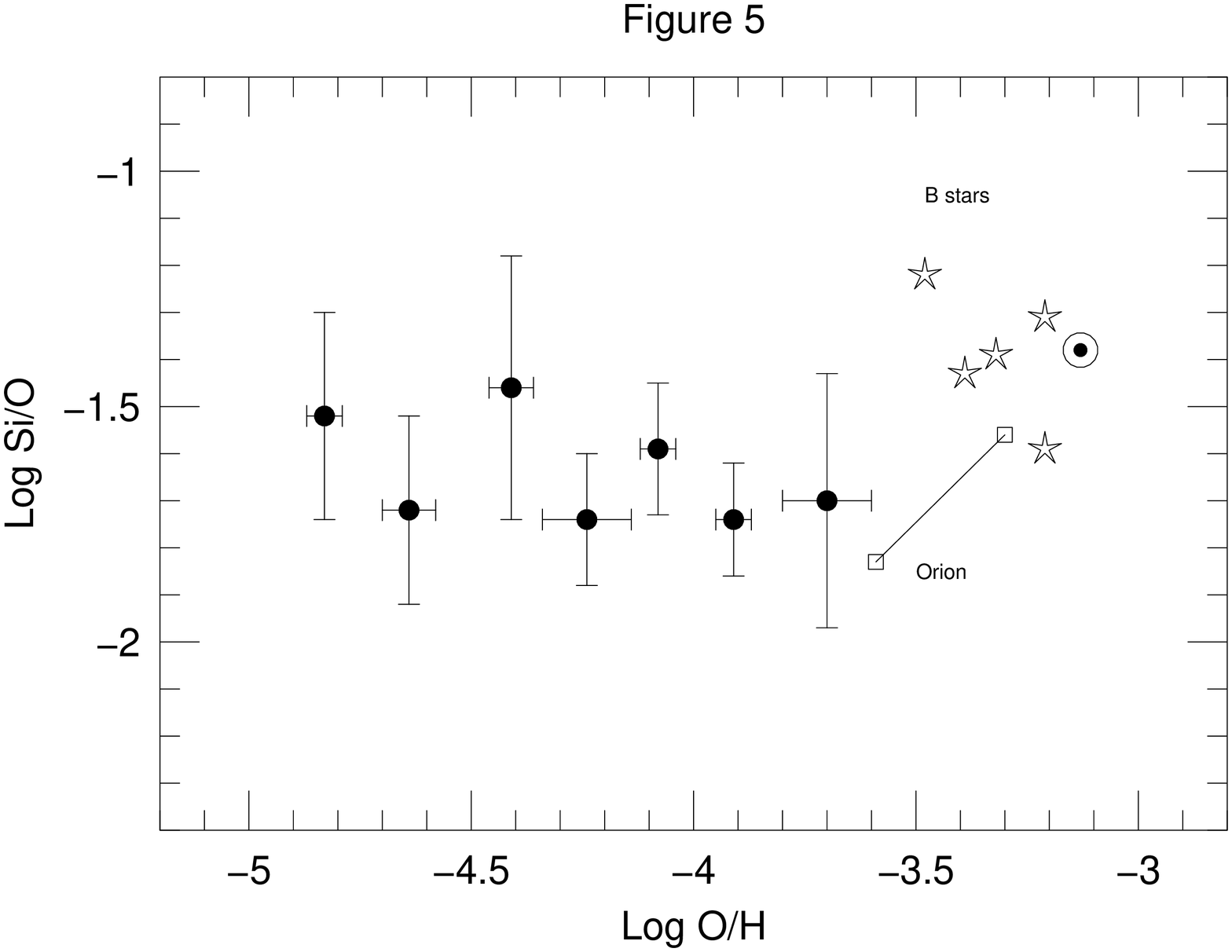}} \hfill
\parbox[b]{\hsize}{
\caption{Si/O abundance ratios in irregular galaxies, from Garnett
et al. 1995b (filled circles). The open squares represent two
different measurements for the Orion nebula, while the stars show
averages for four samples of Galactic B stars (see Garnett et al. 
1995b for details).
}}
\end{figure}

Iron has a variety of emission lines from [Fe~II] and [Fe~III]
in the optical spectrum. The [Fe~III] 4658 \AA\ is often observed
in extragalactic H~II regions. Izotov \& Thuan (1999) measured
[Fe~III] in several metal-poor emission-line galaxies and derived Fe 
abundances. They obtained Fe/O ratios that were similar to the
values found for metal-poor stars in the Galactic halo, and used
this to argue that the their emission-line galaxies were very 
young. However, the ionization corrections for Fe are very uncertain,
and Fe is highly depleted onto grains in the ISM. If Si is depleted
by 0.2 to 0.4 dex, one can use the depletion analysis of Sofia 
et al.  (1994) to estimate that Fe is depleted by about 0.7 to 1.1
dex. It is improbable that O/Fe ratios in H~II regions can be used
to interpret the enrichment history of the ISM without better 
understanding of the depletion factors.

\section {Open Questions and Concluding Remarks}

In conclusion, I'd like to enumerate a few questions regarding the 
chemical evolution of galaxies that seem to need further investigation.

\begin{itemize}
\item
What are the primary mechanisms determining the shape and slope of 
abundance gradients in spiral galaxies? We have seen that chemical
evolution models tend to predict that composition gradients should
get shallower in the inner disks spirals, but this has not been
observed in real galaxies. Does viscous evolution play a role in 
maintaining an exponential gradient, by transferring new gas into 
the inner disk? This is essentially a hydrodynamical problem. It
is also important to determine how closely composition gradients
follow an exponential profile, since the H~II region abundance
scale is still uncertain for high metallicity. IR spectroscopy
will make a significant contribution to constraining the abundance
calibrations in the metal-rich regime.
\item
How homogeneous are abundances in galaxies at a given place and time?
Conflicting studies have argued for significant ($\pm$0.2-0.3 dex) 
variations in abundances on small scales, or for a very homogeneous
composition ($<$ 0.1 dex variations). The question is relevant to
the time scales for cooling and mixing of stellar ejecta with the
ambient ISM. Most chemical evolution models assume instantaneous
mixing, but is this a good approximation? Detailed tudies of abundance
variations across small ($<$ 1 kpc) regions of galaxies will tell us
how homogeneous the ISM composition is.
\item
Is infall of gas presently occuring in galaxies, and how does the
rate evolve with time? This is a big unknown, since most chemical
evolution models use relatively slow, ongoing infall of metal-poor 
gas to suppress the fraction of metal-poor G dwarfs. Observational
evidence for classical infall is sketchy. The high-velocity clouds
seen at high latitudes may represent infall, or may be part of a
Galactic fountain flow, or may be associated with interaction 
between the Galaxy and its satellite galaxies.
\item
What galaxies may be losing metals to the IGM, and how much do
they contribute? Is there a threshold mass above which galaxies
retain metals? 
\item
Does galaxy environment influence composition? The Virgo cluster
studies need to be followed up by larger samples in a wider variety
of cluster environments. The spiral-rich Ursa Major cluster and 
the Coma cluster are two obvious choices for continued study.
\item
Is there zero metallicity gas (or stars for that matter) in galaxies?
Pre-enrichment by an initial stellar Population III also provides
a solution to the G-dwarf problem and to the origin of metals seen
in Ly$\alpha$ forest clouds. What is the composition of the huge gas 
reservoirs in the outer parts of spirals and irregulars?
\end{itemize}

For abundance work in ionized nebulae, we need to understand better
the effects of dust on the thermal and ionization balance. The 
calculated ionizing spectra from massive stars are still in a
state of flux, as more physics and opacity are included; this is
an area that will continue to require attention as computing power
grows. The effects of inhomogeneous structure on the observed 
emission-line spectrum of H~II regions also needs to be addressed.

Stellar nucleosynthesis also needs continuing attention. The biggest 
remaining problem in theoretical stellar evolution and nucleosynthesis 
continues to be the treatment of convective mixing, which affects both
structure and nucleosynthesis. This is also a hydrodynamical problem
requiring improved computing power, and should provide a source of
entertainment (and argument) for some time.

We are seeing great improvements in the study of abundances in nearby 
galaxies, particularly with the new space-based UV and IR observatories,
which are greatly improving the data for elements besides oxygen. With 
these new data, we are in a better position to connect the present-day 
abundance patterns in galaxies with those observed in high-redshift gas 
clouds. Eventually, emission-line spectroscopy of distant galaxies should 
greatly expand our information on heavy element abundances at early times, 
and allow us to trace the evolution of metallicity in the universe in 
greater detail. This will be an important complement to the absorption
line work on DLAs, as the connection between the emission-line gas and
the stellar component is much more clear, and the disk component can
be sampled more completely than with absorption studies.

For nearby galaxies one challenge for observers is to compile a homogeneous 
reference set of abundance data, to provide a statisically significant 
sample for outlining the relationships between abundances, galaxy mass, 
and Hubble type, and for understanding the effects of environment on 
abundance profiles. Some Hubble types are poorly represented in the 
database, particularly very early Hubble types (Sa-Sab), and very late 
types (Sd). Basic structural data for nearby galaxies also need improvement. 
The amount and distribution of molecular gas is one area for improvement.
Stellar mass-to-light ratios and the stellar mass surface density
distribution is another. Wide-field imaging in the infrared should
help reduce the uncertainty in the mass of the stellar component.

One might have noticed that many of the theoretical questions mentioned 
above come down to hydrodynamics. Understanding star formation, the 
evolution of galaxies, and the structure of stars all involve hydrodynamics
at fundamental levels, so I believe that improved hydrodynamical modeling
of all of these phenomena will be the key to a better understanding of
galaxy evolution. Understanding the mechanisms which connect abundances
in galaxies to galaxy structure should provide a continuing challenge 
to galaxy evolution theorists. 

\begin{acknowledgments}
I am grateful to the organizers of this Winter School for the opportunity
to meet and interact with the young scientists (on both the galactic and 
stellar side) whose research papers I have been enjoying in the past
couple of years. Special thanks go to Eric Bell for producing Figure 5
for me, and for informative discussions of the properties of galaxy
colors and population synthesis models. The review presented here has
also benefitted by numerous conversations over the past years with 
Daniela Calzetti, Mike Edmunds, Gary Ferland, Claus Leitherer, John 
Mathis, Andy McWilliam, and Verne Smith. Finally, I must also acknowledge 
my various collaborators on galaxy abundances (Reginald Dufour, Rob 
Kennicutt, Greg Shields, Evan Skillman, Manuel Peimbert, and Silvia 
Torres-Peimbert) who have contributed greatly to many of the results
I have presented in these lectures. My work on abundances in galaxies
has been supported the past four years by NASA grant NAG5-7734.
\end{acknowledgments}

$\;$

\begin{thebibliography}{} 
%
%
\bibitem[]{af62}{\sc Aller, L. H. \& Faulkner, D. J.}, 1962, PASP 74, 219

\bibitem[]{alloin79}{\sc Alloin, D., Collin-Soufrin, S., Joly, M., \& 
Vigroux, L.}, 1979, A\&A 78, 200

\bibitem[]{ag89}{\sc Anders, E. and Grevesse, N.}, 1989, 
Geochim. Cosmochim. Acta 53 197

\bibitem[]{apb95}{\sc Andredakis, Y. C., Peletier, R. F., \& Balcells, M.}
1995, MNRAS 275, 874

\bibitem[]{adc91}{\sc Armandroff, T. E. \& da Costa, G. S.} 1991, 
AJ 100 1329

\bibitem[]{arnett96}{\sc Arnett, D.}, 1996, Supernovae and Nucleosynthesis,
(Princeton University Press) 

\bibitem[]{barnes91}{\sc Barnes, J.}, 1991, IAU Symposium 126, Dynamics of
Galaxies and Their Molecular Cloud Distributions, eds. F. Combes and F.
Casoli (Dordrecht: Kluwer), 363

\bibitem[]{bcf96}{\sc Baugh, C. M., Cole, S., \& Frenk, C. S.}, 
1996, MNRAS 283, 1361

\bibitem[]{bdj00}{\sc Bell, E. F., \& de Jong, R. S.}, 2000, MNRAS 312, 497 

\bibitem[]{bdj01}{\sc Bell, E. F., \& de Jong, R. S.}, 2001, ApJ 550, 212 

\bibitem[]{bos}{\sc Bosch, G., Terlevich, R., Melnick, J., \& Sehman,
F.}, 1999, A\&AS 137, 21 

\bibitem[]{bst95}{\sc Bregman, J. N., Schulman, E., \& Tomisaka, K.}, 
1995, ApJ 439, 155

\bibitem[]{bkg99}{\sc Bresolin, F., Kennicutt, R. C., Jr., \& Garnett, D. R.}, 
1999, ApJ 510, 104

\bibitem[]{br92}{\sc Broeils, A. H.}, 1992, PhD thesis, University of Groningen

\bibitem[]{bvw94}{\sc Broeils, A. H., \& van Woerden, H.}, 1994, A\&AS 107, 129 

\bibitem[]{bfg86}{\sc Burstein, D., Faber, S. M., \& Gonz\'alez, J. J.}, 
1986, AJ 91, 1130

\bibitem[]{cks94}{\sc Calzetti, D., Kinney, A. L., \& Storchi-Bergman, T.} 
1994, ApJ 429, 582

\bibitem[]{car96}{\sc Carigi, L.}, 1996, RMxAA 32, 179

\bibitem[]{car00}{\sc Carigi, L.}, 2000, RMxAA 36, 171

\bibitem[]{cvg91}{\sc Casertano, S., \& van Gorkom, J. H.}, 1991, AJ 101, 1231

\bibitem[]{ckbvg94}{\sc Cayatte, V., Kotanyi, C., Balkowski, C., \& van Gorkom, 
J. H.}, 1994, AJ 107, 1003

\bibitem[]{cla}{\sc Clarke, C. J.} 1989, MNRAS, 238, 283

\bibitem[]{com}{\sc Combes, F.}, 1998, in Abundance Profiles: Diagnostic Tools 
for Galaxy History, eds. D. Friedli, M. Edmunds, C. Robert, and L. Drissen (San 
Francisco: ASP), 300

\bibitem[]{cdfp90}{\sc Combes, F., Debbasch, F., Friedli, D., \& Pfenniger, D.}, 
1990, A\&A 233, 82

\bibitem[]{dekhh97}{\sc de Koter, A., Heap, S. R., \& Hubeny, I.}, 1997, ApJ, 
477, 792

\bibitem[]{dejong96}{\sc de Jong, R. S.} 1996, A\&AS, 118, 557

\bibitem[]{dev:pen}{\sc de Vaucouleurs, G., \& Pence, W. D.}, 1978, AJ, 83, 1163

\bibitem[]{sauron02}{\sc de Zeeuw, P. T., et al.}, 2002, MNRAS, 329, 513

\bibitem[]{ds86}{\sc Dekel, A., \& Silk, J.} 1986, ApJ, 303, 39

\bibitem[]{dtt02}{\sc Denicol\'o, G., Terlevich, R., \& Terlevich, E.} 2002, 
MNRAS 330, 695

\bibitem[]{dp00}{\sc D\'\i az, A. I., \& P\'erez-Montero, E.} 2000, MNRAS 312, 130

\bibitem[]{diaz91}{\sc D\'\i az, A. I., Terlevich, E., V\'\i lchez, J. M.,
Pagel, B. E. J., \& Edmunds, M. G.} 1991, MNRAS 253, 245

\bibitem[]{dressler80}{\sc Dressler, A.} 1980, ApJ, 236, 351

\bibitem[]{dgs88}{\sc 
Dufour, R. J., Garnett, D. R., \& Shields, G. A.,} 1988 ApJ 332, 752

\bibitem[]{ds82}{\sc 
Dufour, R. J., Shields, G. A., \& Talbot, R. J., Jr.}  1982 ApJ 252, 461

\bibitem[]{edm90}{\sc Edmunds, M. G.}, 1990, MNRAS, 246, 678

\bibitem[]{edm:pag}{\sc Edmunds, M. G., \& Pagel, B. E. J.}, 1984, 
MNRAS, 211, 507

\bibitem[]{els62}{\sc Eggen, O., Lynden-Bell, D., \& Sandage, A.}, 1962, 
ApJ, 136, 748

\bibitem[]{essp99}{\sc Ellison, S. L., Songaila, S., Schayes, J., \& 
Pettini, M.}, 1999, AJ, 120, 1175

\bibitem[]{fer98}{\sc Ferland, G. J.}, 1998, Ringberg Workshop on the
Orion Complex (astro-ph/9808107)

\bibitem[]{fc97}{\sc Forestini, M., \& Charbonnel, C.},
1997 A\&AS 123, 241

\bibitem[]{fbk94}{\sc Friedli, D., Benz, W., \& Kennicutt, R. C. Jr.},
1994 ApJ 430, L105

\bibitem[]{gar89}{\sc Garnett, D. R.}, 1989 ApJ 345, 282

\bibitem[]{gar90}{\sc Garnett, D. R.}, 1990 ApJ 363, 142

\bibitem[]{gar92}{\sc Garnett, D. R.}, 1992 AJ 103, 1330

\bibitem[]{gar99}{\sc Garnett, D. R.}, 1999 in Spectrophotometric Dating
of Stars and Galaxies, eds. I. Hubeny, S. Heap, and R. H. Cornett (San
Francisco: ASP), 61

\bibitem[]{gar02}{\sc Garnett, D. R.}, 2002, in preparation

\bibitem[]{gd01}{\sc Garnett, D. R., \& Dinerstein, H. L.}, 
2001, RMxAASC 10, 13

\bibitem[]{gd02}{\sc Garnett, D. R., \& Dinerstein, H. L.}, 
2002, ApJ 558, 145

\bibitem[]{gdp95i}{\sc 
Garnett, D. R., Dufour, R. J., Peimbert, M., Torres-Peimbert, S., 
Shields, G. A., Skillman, E. D., Terlevich, E., \& Terlevich, R. J.,}, 
1995b, ApJ 449, L77

\bibitem[]{gk94}{\sc Garnett, D. R., \& Kennicutt, R. C. Jr.},
1994, ApJ 426, 123

\bibitem[]{gkcs91}{\sc Garnett, D. R., Kennicutt, R. C. Jr., Chu, Y.-H.,
\& Skillman, E. D.} 1991, ApJ 373, 458

\bibitem[]{gar:shi}{\sc Garnett, D. R., \& Shields, G. A.},
1987, ApJ, 317, 82

\bibitem[]{gar:shi2}{\sc 
Garnett, D. R., Shields, G. A., Skillman, E. D., Sagan, S. P., and Dufour, 
R. J.},  1997a, ApJ 489, 63

\bibitem[]{gar:shi3}{\sc 
Garnett, D. R., Shields, G. A., Peimbert, M., Torres-Peimbert, S., Skillman,
E. D., Dufour, R. J., Terlevich, E., and Terlevich, R. J.},
1999, ApJ 513, 168

\bibitem[]{gar:ski}{\sc 
Garnett, D. R., Skillman, E. D., Dufour, R. J., Peimbert, M., Torres-Peimbert, 
S., Terlevich, E., Terlevich, R. J., and Shields, G. A.}, 1995a, ApJ 443, 142

\bibitem[]{gar:ski2}{\sc 
Garnett, D. R., Skillman, E. D., Dufour, R. J., and Shields, G. A.},
1997b, ApJ 481, 174

\bibitem[]{gk92}{\sc 
G\"otz, M., \& K\"oppen, J.}, 1992 A\&A 260, 455

\bibitem[]{gg72}{\sc 
Gunn, J. E., \& Gott, J. R.}, 1972 ApJ 176, 1

\bibitem[]{gus:kar}{\sc 
Gustafsson, B., Karlsson, T., Olsson, E., Edvardsson, B., and Ryde, N.},
1999, A\&A 342, 426

\bibitem[]{hale98}
{\sc Hale, G. M.} 1998, Stellar Evolution, Stellar Explosions, and 
Galactic Chemical Evolution, ed. A. Mezzacappa (Bristol: Institute of
Physics), 17

\bibitem[]{hpn93}
{\sc Hasan, H., Pfenniger, D., \& Norman, C.} 1993, ApJ 409, 91

\bibitem[]{henry89}
{\sc Henry, R. B. C. } 1989, MNRAS 241, 453

\bibitem[]{hbcep96}
{\sc Henry, R. B. C., Balkowski, C., Cayatte, V., Edmunds, M. G., \& 
Pagel, B. E. J.} 1996, MNRAS 293, 635

\bibitem[]{hpg94}
{\sc Henry, R. B. C., Pagel, B. E. J., \& Chincarini, G. L.} 1994, MNRAS 266, 421

\bibitem[]{hw99}
{\sc Henry, R. B. C., \& Worthey, G.} 1999, PASP 111, 919

\bibitem[]{hbol96}
{\sc Hunter D. A., Baum, W. A., O'Neil, E. J., Jr., \& Lynds, R.} 
1996, ApJ 456, 174

\bibitem[]{hunt95}
{\sc Hunter D. A., Shaya, E. J., Holtzman, J. A., Light, R. M., O'Neil,
E. J. Jr., \& Lynds, R.} 1995, ApJ 448, 179

\bibitem[]{igi94}{\sc Ibata, R. A., Gilmore, G., \& Irwin, M. J.}
1994 Nature 370, 194

\bibitem[]{ibe:tru}{\sc Iben, I. Jr., and Truran, J. W. Jr.},
1978 ApJ 220, 980

\bibitem[]{is97a}{\sc Israel, F. P.}, 1997a A\&A 317, 65

\bibitem[]{is97b}{\sc Israel, F. P.}, 1997b A\&A 328, 471

\bibitem[]{it98}{\sc Izotov, Y. I., \& Thuan, T. X.}, 1998 ApJ 500, 188

\bibitem[]{it99}{\sc Izotov, Y. I., \& Thuan, T. X.}, 1999 ApJ 511, 639

\bibitem[]{jma96}{\sc Jablonka, P., Martin, P, \& Arimoto, N.} 1996,
AJ 112, 1415

\bibitem[]{jc99}{\sc Jacoby, G. H., \& Ciardullo, R.} 1999, ApJ 515, 169

\bibitem[]{kau96}{\sc Kauffmann, G.}, 1996, MNRAS, 281, 475

\bibitem[]{ky89}{\sc 
Kenney, J. D. P., \& Young, J. S.}, 1989, ApJ, 344, 171

\bibitem[]{kg96}{\sc 
Kennicutt, R. C., Jr., and Garnett, D. R.}, 1996, ApJ, 456, 504

\bibitem[]{ks96}{\sc Kobulnicky, H. A., \& Skillman, E. D.}, 1996, ApJ, 471, 211

\bibitem[]{ks98}{\sc Kobulnicky, H. A., \& Skillman, E. D.}, 1998, ApJ, 497, 601

\bibitem[]{korn02}{\sc 
Korn, A. J., Keller, S. C., Kaufer, A., Langer, N., Przybilla, N., 
Stahl, O., \& Wolf, B.}, 2002, A\&A, in press (astro-ph/0201453)

\bibitem[]{lan:fli}{\sc Langer, N., Fliegner, J., Heger, A., and Woosley, 
S. E.}, 1997 Nucl. Phys. A621, 457

\bibitem[]{latt98}
{\sc Lattanzio, J. C.} 1998, Stellar Evolution, Stellar Explosions, and 
Galactic Chemical Evolution, ed. A. Mezzacappa (Bristol: Institute of
Physics), 299

\bibitem[]{ldwwgs87}{\sc Lester, D. F., Dinerstein, H. L., Werner, M. W., 
Watson, D. M., Genzel, R. L., \& Storey, J. W. V.}, 1987 ApJ 320, 573

\bibitem[]{lrspt79}{\sc Lequeux, J., Rayo, J. F., Serrano, A., 
Peimbert, M., \& Torres-Peimbert, S.}, 1979 A\&A 80, 155

\bibitem[]{lin:pri}{\sc 
Lin, D. N. C., \& Pringle, J. E.},  1987, ApJ, 320, L87

\bibitem[]{liu95}{\sc 
Liu, X.-W., Storey, P. J., Barlow, M. J., \& Clegg, R. E. S.} 1995, 
MNRAS 272, 369

\bibitem[]{liu00}{\sc 
Liu, X.-W., Storey, P. J., Barlow, M. J., Danziger, I. J., Cohen, M., \& 
Bryce, M.} 2000, MNRAS 312, 585

\bibitem[]{lu:sar}{\sc 
Lu, L., Sargent, W. L. W., and Barlow, T. A.}, 1998 AJ 115, 55

\bibitem[]{mae}{\sc Maeder, A.}, 1992 A\&A 264, 105

\bibitem[]{mal:bla}{\sc Maloney, P., \& Black, J. H.}, 1988, ApJ, 389, 401

\bibitem[]{m2k}{\sc Marigo, P.}, 2001, A\&A, 370, 194

\bibitem[]{mbc96}{\sc Marigo, P., Bressan, A., \& Chiosi, C.}, 1996, A\&A, 313, 545

\bibitem[]{mbc98}{\sc Marigo, P., Bressan, A., \& Chiosi, C.}, 1998, A\&A, 331, 580

\bibitem[]{mr90}{\sc Martin, P. G., \& Rouleau, F.}, 1990, in Extreme Ultraviolet
Astronomy, eds. R. F. Malina and S. Bowyer (Oxford: Pergamon), 341

\bibitem[]{mar:roy}{\sc Martin, P., \& Roy, J.-R.}, 1994, ApJ, 424, 599

\bibitem[]{mh02}{\sc Mart\'\i n-Hern\'andez, N. L., Peeters, E., et al.} 
2002, A\&A, 381, 606

\bibitem[]{msh02}{\sc Martins, F., Schaerer, D., \& Hiller, D. J.} 
2002, A\&A 382, 999

\bibitem[]{mateo98}{\sc Mateo, M.}, 1998, ARAA 36, 435

\bibitem[]{math62}{\sc Mathis, J. S.}, 1962, ApJ 136, 374

\bibitem[]{math96}{\sc Mathis, J. S.}, 1996, ApJ 472, 643

\bibitem[]{mat:chi}{\sc 
Matteucci, F., and Chiosi, C.}, 1983 MNRAS 239, 885

\bibitem[]{mat:fra}{\sc 
Matteucci, F., and Fran\c cois, P.}, 1989 MNRAS 239, 885

\bibitem[]{mcc}{\sc McCall, M. L.}, 1982, PhD thesis, University of Texas at Austin

\bibitem[]{mlvkpn95}{\sc McCarthy, J. K., Lennon, D. J., Venn, K. A., 
Kudritzki, R.-P., Puls, J., \& Najarro, F.}, 1995, ApJ 455, L135

\bibitem[]{mcwr94}{\sc McWilliam, A. \& Rich, R. M.}, 1994, ApJS 91, 749

\bibitem[]{mb99}{\sc Mighell, K. J. \& Burke, C. J.}, 1999, AJ 118, 366

\bibitem[]{mih}{\sc Mihos, J. C.}, 2001, in {\sc Galaxy Disks and Disk Galaxies},
ASP Conference Series Vol. 230, eds. J. G. Funes and E. M. Corsini, p. 491

\bibitem[]{mmw98}{\sc 
Mo, H., Mao, S., \& White, S. D. M.}, 1998, MNRAS, 295, 319

\bibitem[]{mfd97}{\sc 
Moll\'a, M., Ferrini, F., \& D\'\i az, A. I.}, 1997, ApJ, 475, 519

\bibitem[]{mhlk97}{\sc 
Monteverde, M. I., Herrero, A., Lennon, D. J., \& Kudritzki, R.-P.}, 
1997, ApJ, 474, 107

\bibitem[]{ok93}{\sc Oey, M. S., \& Kennicutt, R. C. Jr.} 1993 ApJ 411, 137

\bibitem[]{ost89}{\sc Osterbrock, D. E.} 1989, Astrophysics of Gaseous
Nebulae and Active Galactic Nuclei (Mills Valley, CA: University Science Books)

\bibitem[]{oss97}{\sc Olive, K. A., Skillman, E. D., \& Steigman, G.}, 
1997 ApJ 483, 788

\bibitem[]{pebcs79}{\sc 
Pagel, B. E. J., Edmunds, M. G., Blackwell, D. E., Chun, M. S., \& Smith, G.}, 
1979, MNRAS 189, 95

\bibitem[]{pa86}{\sc 
Pakull, M. W., \& Angebault, L. P.}, 1986, Nature 322, 511

\bibitem[]{pptp86}{\sc Peimbert, M., Pe\~na, M., \& Torres-Peimbert, S.}, 
1986 A\&A 158, 266

\bibitem[]{ps70}{\sc Peimbert, M., \& Spinrad, H.}, 1970 ApJ 159, 809

\bibitem[]{ptp74}{\sc Peimbert, M., \& Torres-Peimbert, S.}, 1974 ApJ 193, 327

\bibitem[]{pel99}{\sc 
Pelletier, R. F., et al.} 1999, MNRAS, 310, 863

\bibitem[]{pet:lip}{\sc 
Pettini, M., Lipman, K., and Hunstead, R. W.}, 1995, ApJ, 451, 100

\bibitem[]{phi:edm}{\sc 
Phillips, S., \& Edmunds, M. G.}, 1991, MNRAS, 251, 84

\bibitem[]{pcb98}{\sc 
Portinari, L., Chiosi, C., \& Bressan, A.}, 1998, A\&A 334, 505

\bibitem[]{pr:bo}{\sc 
Prantzos, N., \& Boissier, S.}, 2000, MNRAS, 313, 338

\bibitem[]{pvc94}{\sc 
Prantzos, N., Vangioni-Flam, R., \& Chauveau, S.}, 1994, A\&A, 285, 132

\bibitem[]{rv81}{\sc 
Renzini, A., \& Voli, M.}, 1981, A\&A 94, 175

\bibitem[]{rmcw00}{\sc 
Rich, R. M., \& McWilliam, A.}, 2000, Proc. SPIE 4005, 150

\bibitem[]{rsdr00}{\sc 
Rolleston, W. R. J., Smartt, S. J., Dufton, P. L., \& Ryans, R. S. I.}, 
2000, A\&A 363, 537

\bibitem[]{rw97}{\sc 
Roy, J.-R.., \& Walsh, J. R.}, 1997, MNRAS, 288, 715

\bibitem[]{rr98}{\sc 
Rudnick, G., \& Rix, H.-W.}, 1998, AJ, 116, 1163

\bibitem[]{ryd}{\sc 
Ryder, S. D.}, 1995, ApJ, 444, 610

\bibitem[]{s71}{\sc 
Searle, L.}, 1971, ApJ, 168, 327

\bibitem[]{ss72}{\sc 
Searle, L., \& Sargent, W. L. W.}, 1972, ApJ, 173, 25

\bibitem[]{sz78}{\sc 
Searle, L., \& Zinn, R.}, 1978, ApJ, 225, 357

\bibitem[]{she:cot}{\sc 
Shetrone, M. D., Cot\'e, P., \& Sargent, W. L. W.},  
2001, ApJ, 548, 592

\bibitem[]{sk95}{\sc 
Shields, J. C., \& Kennicutt, R. C. Jr.}, 1995, ApJ, 454, 807

\bibitem[]{skill98}{\sc 
Skillman, E. D.}, 1998, Stellar Astrophysics for the Local Group, eds.
A. Aparicio, A. Herrero, and F. S\'anchez (Cambridge University Press), 457

\bibitem[]{ski:ben}{\sc 
Skillman, E. D., \& Bender, R.}, 1995, RMxAASC, 3, 25

\bibitem[]{ski:ken}{\sc 
Skillman, E. D., Kennicutt, R. C., Jr., Shields, G. A., \& Zaritsky, D.},  
1996, ApJ, 462, 147

\bibitem[]{shshl96}{\sc 
Smecker-Hane, T. A., Stetson, P. B., Hesser, J. E., \& Lehnert, M. D.}, 
1994 AJ 108, 507

\bibitem[]{sne:cow}{\sc 
Sneden, C., Cowan, J. J, Ivans, I. I., Fuller, G. M., Burles, S., Beers, T. C.,
\& Lawler, J. E.}, 2000 ApJ 533, L139

\bibitem[]{sof:car}{\sc 
Sofia, U. J., Cardelli, J. A, Guerin, K. P., and Meyer, D. M.}, 1997 ApJ 
482, L105

\bibitem[]{scs94}{\sc 
Sofia, U. J., Cardelli, J. A, \& Savage, B. D.}, 1994 ApJ 430, 650

\bibitem[]{stsz01}{\sc 
Stasi\'nska, G., \& Szczerba, G. P.}, 2001, A\&A 379, 1024

\bibitem[]{tay:gup}{\sc 
Tayal, S. S., \& Gupta, G. P.}, 1999 ApJ 526, 544

\bibitem[]{thu:izo}{\sc 
Thuan, T. X., Izotov, Y. I., and Lipovetsky, V. A.}, 1995 ApJ 445, 108

\bibitem[]{tinsl79}{\sc 
Tinsley, B. M.}, 1979 ApJ 229, 1046

\bibitem[]{tols01}{\sc 
Tolstoy, E., Irwin, M. J., Cole, A. A., Pasquini, L., Gilmozzi, R., 
\& Gallagher, J. S.}, 2001 MNRAS 327, 918

\bibitem[]{tor:pei}{\sc 
Torres-Peimbert, S., Peimbert, M. and Fierro, J.}, 1989 ApJ 345, 186

\bibitem[]{tfwg00a}{\sc 
Trager, S. C., Faber, S. M., Worthey, G., \& Gonz\'alez, J. J.}, 2000a, 
AJ 119, 1645

\bibitem[]{tfwg00b}{\sc 
Trager, S. C., Faber, S. M., Worthey, G., \& Gonz\'alez, J. J.}, 2000b, 
AJ 120, 165

\bibitem[]{tsu:yos}{\sc 
Tsujimoto, T., Yoshii, Y., Nomoto, K. and Shigeyama, T.}, 1995 A\&A 302, 704

\bibitem[]{tyson86}{\sc 
Tyson, J. A.}, 1986 J. Opt. Soc. Amer. A 3, 2131

\bibitem[]{vdhg97}{\sc 
van den Hoek, L. B., \& Groenewegen, M. A. T.}, 1997 
A\&AS 123, 305

\bibitem[]{vanzee98}{\sc 
van Zee, L., Salzer, J., Haynes, M., O'Donoghue, A. \& Balonek, T.}, 1998 
AJ 116, 2805

\bibitem[]{vil:edm}{\sc 
Vila-Costas, M. B., \& Edmunds, M. G.},  1992, MNRAS, 259, 121 (VCE)

\bibitem[]{vcbh01}{\sc 
Vladilo, G., Centuri\'on, M., Bonifacio, P. \& Howk, J. C.}, 2001 ApJ 557, 1007

\bibitem[]{walb91}{\sc Walborn, N. R.} 1991, IAU Symposium 148: The Magellanic
Clouds, eds. R. Haynes and D. Milne (Dordrecht: Kluwers), 145

\bibitem[]{warme86}{\sc Warmels, R. H.} 1988, A\&AS 72, 427 

\bibitem[]{wea:woo}{\sc 
Weaver, T. A., \& Woosley, S. E.}, 1993 Phys. Rep. 227, 65 (WW93)

\bibitem[]{whi99}{\sc 
Whiting, A. B., Hau, G. K. T., \& Irwin, M. J.}, 1999 AJ 118, 2767

\bibitem[]{wils95}{\sc Wilson, C. D.} 1995 ApJ 448, L97 

\bibitem[]{worth98}{\sc Worthey, G.} 1998 PASP 110, 888 

\bibitem[]{yun94}{\sc 
Yun, M. S., Ho, P. T. P., \&  Lo, K. Y.}, 1994 Naturee 372, 530

\bibitem[]{zar95}{\sc Zaritsky, D.}, 1995, ApJ, 448, L17

\bibitem[]{zar:ken}{\sc 
Zaritsky, D., Kennicutt, R. C. Jr., \& Huchra, J. P.},  1994, ApJ, 420, 
87 (ZKH)

\end{thebibliography}
\end{document}